# A Numerical Approach to Non-Equilibrium Quantum Thermodynamics: Non-Perturbative Treatment of the Driven Resonant Level Model based on the Driven Liouville von-Neumann Formalism

Annabelle Oz,[1,2] Oded Hod,[1,2] and Abraham Nitzan[1,2,3]

[1] *Department of Physical Chemistry, School of Chemistry, The Raymond and Beverly Sackler Faulty of Exact Sciences, Tel Aviv University, Tel Aviv, IL 6997801*

[2] *The Sackler Center for Computational Molecular and Materials Science, Tel Aviv University, Tel Aviv, IL 6997801*

[3] *Department of Chemistry, University of Pennsylvania, Philadelphia, PA, USA 19103*

## Abstract

Non-equilibrium thermodynamics of the driven resonant level model is studied using numerical simulations based on the driven Liouville von-Neumann formalism. The approach is first validated against recently obtained analytical results for quasi-static level shifts and the corresponding first order corrections. The numerical approach is then used to study far-from-equilibrium thermodynamic properties of the system under finite level shift rates. The proposed methodology allows the study of unexplored non-equilibrium thermodynamic regimes in open quantum systems.

# Introduction

The quantum technology revolution promises unprecedented advances in our computation and technological capabilities (see Ref. [1] and references therein). As machines are scaled-down to the quantum regime, it is of prime importance to understand how quantum effects are manifested in their operation. Within this effort, an extension of the thermodynamic description to nanoscale systems out of equilibrium, where dynamic quantum effects dominate, constitutes an important challenge [2,3].

A generic problem of this kind is the thermodynamic analysis of a small quantum system that is coupled to multiple reservoirs, which are out of equilibrium with respect to each other, possibly with an additional external time-dependent force that performs work on the system [2–9]. As in standard macroscopic thermodynamics, such an analysis requires the partitioning of energetic variations in the system into heat and useful work components. However, the small system size makes the characterization of its thermodynamic properties uncertain because energy parameters associated with the system are of the same order as those characterizing the system-baths interactions. Furthermore, quantum mechanics implies that energy transfer and relaxation are associated with broadening of energy levels, so even energy observables of the system alone are not well characterized.

Models constructed to consider these issues are often studied in the weak system-bath coupling limit, where the thermodynamic functions of the system are well defined and it is possible to work with a standard quantum master equation (see Ref. [2] and references therein). Studies that consider the implications of strong system-bath coupling have recently emerged [4–6,10–15]. In the latter, the use of weak system-reservoir(s) coupling is replaced by the assumption that the dynamics imposed on the system is slow. "Slow" here implies that the timescale on which the system Hamiltonian is changed by external force(s) (as work is done or extracted from the system) is long relative to the characteristic timescales on which the system exchanges energy and particles with its environment. We also note that these treatments are usually limited to non-interacting systems (e.g., free-electron or Harmonic bath models), although extensions of such considerations to include electron-electron interactions in the bath have recently been published [15,16].

While such studies provide fundamental insights into dynamics and thermodynamics of small systems in these specific limits, it is obviously of interest to explore less restrictive conditions, e.g. strong coupling under arbitrary external driving. To this end, one may resort to numerical simulations [15,17–28]. A convenient numerical framework, relevant to problems where a small electronic system is coupled to one or more free electron reservoirs, each in its own equilibrium but not necessarily in equilibrium with each other, is the Driven Liouville von Neumann (DLvN) method [29–32,35]. In this approach, the Liouville von-Neumann (LvN) equation of motion (EOM) for an extended system comprising the system of interest plus finite lead models is solved with open boundary conditions imposed on the leads. These boundary conditions are enforced by augmenting the LvN equations with non-unitary source/sink terms constructed to drive each lead towards its own equilibrium state. When the driving is done by imposing different equilibrium states on different leads, the DLvN method was shown to reproduce the results of explicit time-dependent non-equilibrium Green's function treatments [33], while avoiding violations of Pauli's exclusion principle and preserving density matrix positivity - two issues that were encountered in previous related implementations [25,34]. This was formally supported by a derivation of the DLvN EOM as an approximation resulting from the non-equilibrium Green's function formalism [33,35] and its recasting in Lindblad form for non-interacting systems [33,37].

The purpose of the present work is to examine the applicability of the DLvN methodology for non-equilibrium thermodynamics simulations. For simplicity and clarity, we limit ourselves to the driven resonant level model considered in Refs. [4,5,7,10,13,38], represented by a time-dependent Hamiltonian comprising of a single electronic level (termed as "dot") coupled to a single equilibrium reservoir and driven by an external force that performs work on the system or extracts work from it. Some technical considerations relevant to the implementation of the DLvN method for evaluating thermodynamic properties of this model system were recently discussed [39]. Here, we use the method to calculate the dynamic evolution of this system numerically and investigate the thermodynamic implications of the driving without being limited to the weak coupling and/or slow driving limits considered in previous studies.

## 2. Methodology

**The Resonant Level Model System**

As mentioned above, for our model system we choose the resonant level model previously considered in Refs. [4,5,7,10,13,38]. The model consists of a single spin-less electronic level that represents a quantum dot, coupled to a spin-less free-electron reservoir representing a metallic lead. The dot level is driven (e.g. by an external gate voltage or an electromagnetic field) such that its energy (and consequently its occupation) varies with time at a finite rate (see Figure 1). The Hamiltonian describing this system is:

$$H = H_0 + V = H_d + H_L + V, \quad (1)$$

where the dot and lead Hamiltonians, $H_d$ and $H_L$, and the coupling $V$ between them are given by:

$$H_d = \varepsilon_d(t) c_d^\dagger c_d, \quad (2)$$

$$H_L = \sum_l \varepsilon_l c_l^\dagger c_l, \quad (3)$$

and

$$V = \sum_l (v_l c_d^\dagger c_l + h.c.). \quad (4)$$

In Eqs. (2)-(4) $c_d^\dagger (c_d)$ and $c_l^\dagger (c_l)$ are the creation (annihilation) operators for an electron in the dot and lead levels of energies $\varepsilon_d$ and $\varepsilon_l$, respectively. For simplicity, the lead level energies and the dot/lead coupling matrix elements, $v_l$, are kept real and constant during the dynamics. Furthermore, the lead is assumed to be at (or sufficiently close to) thermal equilibrium, characterized by electronic temperature $T$ and chemical potential $\mu$. Without loss of generality, the latter is set to be at the energy origin such that $\mu = 0$, as discussed below.

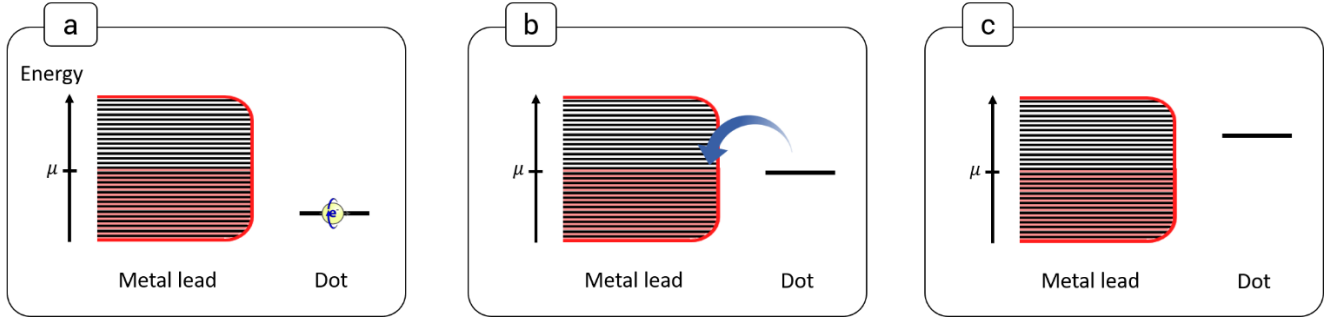

Figure 1: Illustration of the model system. (a) The dot is initially placed well below the Fermi level and is (nearly) fully occupied. (b) As the dot's energy is shifted up, it empties into the lead. (c) When located well above the Fermi level, the dot is (nearly) unoccupied.

Before considering far-from-equilibrium scenarios, we first demonstrate the performance of our numerical calculations at near-equilibrium conditions, where analytical results can be obtained from standard equilibrium quantum statistical mechanics [4]. As will become obvious below, the DLvN numerical scheme is not limited to some approximations invoked in analytical derivations, such as the wide band limit (WBL) and slow driving rates with respect to typical relaxation times in the system. On the other hand, our numerical scheme can only address finite systems, which here implies a free-electron lead modeled by a finite number of levels spanned within a finite bandwidth. Still, to facilitate comparison with the analytical results, we construct our model to be a good representation of the WB approximation by representing the lead by a single energy band of width $W$ and a finite number, $N_L$, of equally spaced (spacing of $\Delta\varepsilon_l = W/N_L$ and denisty of states $\rho = \Delta\varepsilon_l^{-1}$) single electron levels and an energy independent lead-dot coupling, $v$. The spectral function of the dot state is then given, to a good approximation, by a Lorentzian centered around $\varepsilon_d$:

$$A(\varepsilon; \varepsilon_d, \gamma) = \frac{\hbar\gamma}{(\varepsilon-\varepsilon_d)^2 + \left(\frac{\hbar\gamma}{2}\right)^2} \ ; \ \gamma \equiv \frac{2\pi}{\hbar}|v|^2 \Delta\varepsilon_l^{-1} \tag{5}$$

provided that $\hbar\gamma \gg \Delta\varepsilon_l$, $W \gg \hbar\gamma$ and $\varepsilon_d$ is far (relative to the Lorentzian width, $\hbar\gamma$) from the band edges. For brevity of notation we use below $A(\varepsilon)$ to denote $A(\varepsilon; \varepsilon_d, \gamma)$. When these conditions are not fully satisfied, proper corrections can be applied as described in Ref. [39]. It should be emphasized that deviations from the WB limit do not invalidate the thermodynamic calculations described below, only their comparison with the analytical WBL results.

**The Driven Liouville-von-Neuman Approach**

In the realm of the non-interacting spinless resonant level model described above, all observables can be obtained from the 1-electron density matrix $\sigma_{ij}(t) = \langle c_i(t) c_j^\dagger(t) \rangle$, where $c_i^\dagger(t)$ and $c_i(t)$ are the Heisenberg representations of the electron creation and annihilation operators of state $i$ and the average is taken over the initial many-electron wavefunction. For the present treatment $\boldsymbol{\sigma}$ may be conveniently expressed in the basis of eigenstates of the isolated dot and lead sections:

$$\boldsymbol{\sigma} = \begin{pmatrix} \sigma_d & \boldsymbol{\sigma}_{d,L} \\ \boldsymbol{\sigma}_{L,d} & \boldsymbol{\sigma}_L \end{pmatrix}, \tag{6}$$

where $\sigma_d$ is the dot population, $\boldsymbol{\sigma}_L$ is the density matrix block of the lead, and $\boldsymbol{\sigma}_{dL} = \boldsymbol{\sigma}_{Ld}^\dagger$ are the dot/lead coherences vectors. Correspondingly, the system Hamiltonian matrix representation is represented in the same basis as:

$$\boldsymbol{H} = \begin{pmatrix} \varepsilon_d(t) & \boldsymbol{v}^\dagger \\ \boldsymbol{v} & \boldsymbol{\varepsilon}_L \end{pmatrix}, \tag{7}$$

where $\boldsymbol{\varepsilon}_L$ is a diagonal block of lead level energies, $\varepsilon_l$, that, as mentioned above, are taken to be uniformly distributed over the bandwidth, and $\boldsymbol{v}$ is a column vector containing the identical coupling matrix elements that are related to a given dot-level broadening value of $\gamma$ through Fermi's golden rule of Eq. (5).

As discussed above, a numerical evaluation of the dynamics described by the Hamiltonian (1)-(4) necessarily needs to employ a finite lead model, comprising a finite number of states spanned within a finite energy band, $W$. The time evolution of the single-particle density matrix in such a closed system is given by the standard single-particle LvN EOM: $\frac{d\sigma}{dt} = -\frac{i}{\hbar}[H, \sigma]$. In an out-of-equilibrium simulation, this dynamics results in Poincaré recurrences with a period determined by the energy spacing in the lead, $\Delta\varepsilon_l = W/N_L$ [16,17,22,26,40,41]. Therefore, the closed system model can mimic the behavior of its open counterpart only for short periods limited by the typical reflection time of the electronic wavepacket from the finite system boundaries [39,42]. To overcome this problem we adopt the recently developed DLvN approach for simulating time-dependent electronic transport in open quantum systems [26,30–34]. Within this approach, open system boundary conditions are imposed by augmenting the LvN evolution with additional rate processes (see Eq. (9)) that guide the lead section towards its equilibrium form. The latter is represented by a diagonal matrix $\boldsymbol{\sigma}_L^{eq}$ with populations given by the Fermi-Dirac distribution of the corresponding bath, to which the lead is implicitly coupled:

$$f(\varepsilon; \beta, \mu) = \left[e^{\beta(\varepsilon-\mu)} + 1\right]^{-1}, \tag{8}$$

where $\mu$ is the chemical potential of the implicit bath, $\beta = (k_B T)^{-1}$, $k_B$ is Boltzmann's constant, and $T$ is the bath's electronic temperature. For brevity of notation we use below $f(\varepsilon)$ to denote $f(\varepsilon; \beta, \mu)$. The DLvN EOM for the resonant level model considered herein is of the following form:

$$\frac{d\boldsymbol{\sigma}}{dt} = -\frac{i}{\hbar}[\boldsymbol{H}, \boldsymbol{\sigma}] - \Gamma \begin{pmatrix} 0 & \frac{1}{2}\boldsymbol{\sigma}_{d,L} \\ \frac{1}{2}\boldsymbol{\sigma}_{L,d} & \boldsymbol{\sigma}_L - \boldsymbol{\sigma}_L^{eq} \end{pmatrix}, \tag{9}$$

where

$$\left(\boldsymbol{\sigma}_L^{eq}\right)_{ll'} = \delta_{ll'} f(\varepsilon_l). \tag{10}$$

The lead driving rate, $\Gamma$, represents the timescale on which thermal relaxation takes place in the lead due to its coupling to the implicit bath, which is generally assumed to be fast relative to all other processes of interest. While a physically motivated value for $\Gamma$ can be extracted from the electronic properties of the explicit bath [34], in the present study we set $\hbar\Gamma$, the lead levels broadening due to their coupling to the implicit bath, to be uniform and of the order of the lead level spacing, $\hbar\Gamma \sim \Delta\varepsilon_l$. We have verified that the results presented below are fairly insensitive to the specific choice of lead driving rate in this range (see section 1 of the Supporting Information and relevant discussion in Refs. [37,43]), to the bandwidth of the finite lead model (see section 2 of the Supporting Information), and to the density of lead states (see section 3 of the Supporting Information).

## 3. Thermodynamic Functions and Fluxes

We now turn to describe our approach for evaluating thermodynamic functions and fluxes using the numerical scheme detailed above. We distinguish between quasi-static or sudden jump processes, where thermodynamic functions can be obtained from equilibrium calculations, and finite-rate processes, where kinetic equations are required.

**The Quasi-Static Limit**

In a quasi-static process, the driving parameter(s) (here $\varepsilon_d$) is changed slowly (reversibly) relative to system relaxation processes so that the system can be assumed to remain at equilibrium, characterized by the equilibrium density matrix $\boldsymbol{\sigma}^{eq}(\varepsilon_d)$ of the entire dot-lead system, throughout the process. The equilibrium value of a system observable that corresponds to an operator $\boldsymbol{O}$ is given by:

$$\langle \boldsymbol{O} \rangle^{(0)}(\varepsilon_d) = \text{Tr}(\boldsymbol{\sigma}^{eq}(\varepsilon_d)\boldsymbol{O}), \tag{11}$$

and its time-dependence (used in section 4 below for the evaluation of thermodynamic quantities in the quasi-static limit) is derived from the variation of $\varepsilon_d$, such that:

$$\langle \dot{\boldsymbol{O}} \rangle^{(1)}(\varepsilon_d) = \dot{\varepsilon}_d \frac{d\langle \boldsymbol{O} \rangle}{d\varepsilon_d}. \tag{12}$$

The superscripts in Eqs. (11) and (12) denote the order in $\dot{\varepsilon}_d$ of the calculated quantity. Specifically, the WBL expressions for $\varepsilon_d$-dependent contributions to the particle number, energy, and entropy are given by (See Ref. [4]):

$$N_d^{(0),WBL}(\varepsilon_d) = \int_{-\infty}^{\infty} \frac{d\varepsilon}{2\pi} A(\varepsilon) f(\varepsilon), \tag{13}$$

$$E_d^{(0),WBL}(\varepsilon_d) = \int_{-\infty}^{\infty} \frac{d\varepsilon}{2\pi} \varepsilon A(\varepsilon) f(\varepsilon). \tag{14}$$

$$S_d^{(0),WBL}(\varepsilon_d) = k_B \int_{-\infty}^{\infty} \frac{d\varepsilon}{2\pi} A(\varepsilon) \{f(\varepsilon) ln[f(\varepsilon)] + [1 - f(\varepsilon)] ln[1 - f(\varepsilon)]\} \tag{15}$$

The corresponding quasistatic fluxes of particles, energy, work, heat, and entropy are given by ([4]):

$$\dot{N}^{(1),WBL} = \frac{d\varepsilon_d}{dt} \int_{-\infty}^{\infty} \frac{d\varepsilon}{2\pi} \frac{dA(\varepsilon)}{d\varepsilon_d} f(\varepsilon), \tag{16}$$

$$\dot{E}^{(1),WBL} = \frac{d\varepsilon_d}{dt}\int_{-\infty}^{\infty}\frac{d\varepsilon}{2\pi}\varepsilon\frac{dA(\varepsilon)}{d\varepsilon_d}f(\varepsilon) = \dot{\varepsilon}_d N_d^{(0),an}(\varepsilon_d) + \dot{\varepsilon}_d \int_{-\infty}^{\infty}\frac{d\varepsilon}{2\pi}\varepsilon A(\varepsilon)\frac{df(\varepsilon)}{d\varepsilon}, \tag{17}$$

$$\dot{W}^{(1),WBL} = \dot{\varepsilon}_d N_d^{(0)}(\varepsilon_d), \tag{18}$$

$$\dot{Q}^{(1),WBL} = -T\dot{S}_d^{(1)} = \dot{\varepsilon}_d \int_{-\infty}^{\infty}\frac{d\varepsilon}{2\pi}(\varepsilon - \mu)A(\varepsilon)\frac{df(\varepsilon)}{d\varepsilon}. \tag{19}$$

In the numerical finite-bandwidth model calculation, the equilibrium density matrix, $\boldsymbol{\sigma}^{eq}(\varepsilon_d)$ can be obtained using two different approaches. In one, for any instantaneous $\varepsilon_d$ the Hamiltonian (7) is diagonalized and the equilibrium density matrix of the dot-lead system in the eigenstate representation, $\{|j\rangle\}$, $[\boldsymbol{\sigma}^{eq}(\varepsilon_d)]_{j,j'} = \delta_{j,j'}f(\varepsilon_j)$, is used to calculate the equilibrium expectation value of any single-electron operator. For the thermodynamic functions of interest we get:

$$N^{(0)}(\varepsilon_d) = \sum_j f(\varepsilon_j), \tag{20}$$

$$E^{(0)}(\varepsilon_d) = \sum_j \varepsilon_j f(\varepsilon_j), \tag{21}$$

$$S^{(0)}(\varepsilon_d) = -k_B\sum_j\{f(\varepsilon_j)ln[f(\varepsilon_j)] + [1-f(\varepsilon_j)]ln[1-f(\varepsilon_j)]\}, \tag{22}$$

To obtain $\varepsilon_d$-dependent expressions equivalent to Eqs. (13)-(15) these functions need to be projected onto the dot section as follows:

$$N_d^{(0)}(\varepsilon_d) = \sum_j |\langle d|j\rangle|^2 f(\varepsilon_j), \tag{23}$$

$$E_d^{(0)}(\varepsilon_d) = \sum_j |\langle d|j\rangle|^2 \varepsilon_j f(\varepsilon_j), \tag{24}$$

$$S_d^{(0)}(\varepsilon_d) = -k_B\sum_j|\langle d|j\rangle|^2\{f(\varepsilon_j)ln[f(\varepsilon_j)] + [1-f(\varepsilon_j)]ln[1-f(\varepsilon_j)]\}. \tag{25}$$

The work done in the quasistatic process is obtained as an integral over the population:

$$W^{(0)} = \int d\varepsilon_d N_d^{(0)}(\varepsilon_d), \tag{26}$$

and the corresponding heat can be obtained from the first law:

$$Q^{(0)} = \Delta E^{(0)} - W^{(0)} - \mu\Delta N^{(0)}, \tag{27}$$

Alternatively, an approximate equilibrium density matrix may be calculated as the solution of a Sylvester equation [26,39] obtained by setting $\dot{\boldsymbol{\sigma}} = \boldsymbol{0}$ in Eq. (9) for any given value of $\varepsilon_d$. This can then be used to evaluate all needed expectation values of any single-electron observable $\langle A \rangle = \text{Tr}(\boldsymbol{\sigma A})$. The approximate nature of the latter stems from the fact that we impose thermal equilibrium only on the lead states and not on the eigenstates of the full dot-lead system. This would not matter if the lead was infinite, but small differences are expected when using finite lead models. Comparing the two solutions provides a measure of the suitability of the chosen finite lead model as an approximation for a wide-band lead [39] and can guide our effort to balance between the required accuracy and the computational cost. Within this approach, the occupation of the dot, which can be substituted $N_d^{(0)}$ in expressions (26) and (27), is calculated using:

$$n_d = \sigma_{dd}. \tag{28}$$

In either case, the corresponding quasistatic fluxes are obtained as products of $\dot{\varepsilon}_d$ and the numerical derivatives with respect to $\varepsilon_d$ of the above functions.

**The Sudden Jump Limit**

On the other extreme limit, the system starts at equilibrium with $\varepsilon_d = \varepsilon_{d1}$ and at a given time, marked as $t = 0$, the dot energy makes a sudden jump to $\varepsilon_d = \varepsilon_{d2}$ without changing its population, after which the system is let to relax to the new equilibrium associated with $\varepsilon_{d2}$ in place. Obviously, the work, which is performed only during the sudden jump, is given by:

$$W^{sudden}(\varepsilon_{d1} \to \varepsilon_{d2}) = N_d^{(0)}(\varepsilon_{d1})(\varepsilon_{d2} - \varepsilon_{d1}). \tag{29}$$

In the subsequent relaxation to equilibrium, the dot-lead system exchanges heat and particles with the external bath while no further work is done. The change in thermodynamic state functions is given by $\Delta F = F^{(0)}(\varepsilon_{d2}) - F^{(0)}(\varepsilon_{d1})$, $(F = N, E, S)$ and the heat exchanged with the bath during the relaxation processes can be written in terms of the eigenstates and eigenenergies, $(|i\rangle, \varepsilon_i)$ and $(|j\rangle, \varepsilon_j)$, of the initial $(\varepsilon_d = \varepsilon_{d1})$ and final $(\varepsilon_d = \varepsilon_{d2})$ Hamiltonians of the dot-lead system, respectively. This leads to:

$$Q^{sudden}(\varepsilon_{d1} \to \varepsilon_{d2}) = \sum_j \varepsilon_j \left(n_j - f(\varepsilon_j)\right) - \mu \Delta N, \tag{30}$$

where $n_j = \sum_i |\langle i | j \rangle|^2 f(\varepsilon_i)$ is the initial population of state $j$ after the jump and $\Delta N = \sum_j \left(n_j - f(\varepsilon_j)\right)$ is the total change in particle number occurring during the relaxation stage. It is easy to realize that the first law, $Q^{sudden} = \Delta E - W^{sudden} - \mu \Delta N$ is satisfied by writing the energy flux, work per unit time (power), and heat flux in the following forms:

$$\dot{E} = \text{Tr}(\dot{\boldsymbol{\sigma}} \boldsymbol{H}) + \text{Tr}(\boldsymbol{\sigma} \dot{\boldsymbol{H}}) \tag{31}$$

$$\dot{W} = \text{Tr}(\boldsymbol{\sigma}\dot{\boldsymbol{H}}) \tag{32}$$

$$\dot{Q} = \text{Tr}(\dot{\boldsymbol{\sigma}}(\boldsymbol{H} - \mu \boldsymbol{I})), \tag{33}$$

where $\boldsymbol{I}$ is a unit matrix of appropriate dimensions. Clearly, more heat is generated by the thermalization of electrons following the sudden jump than in the quasistatic process, hence $Q^{sudden} > T\Delta S$.

**Finite Dot-Level Driving Rate**

When the dot energy is shifted at a finite rate, the system goes out of equilibrium and irreversibility effects are manifested. Analytical results for the changes in thermodynamic functions and their fluxes can be obtained by expansions with respect to the driving speed $\dot{\varepsilon}_d$, whereupon the lowest order corrections in the WBL are obtained in the forms [4]:

$$N_d^{(1),WBL}(\varepsilon_d) = -\frac{\hbar\dot{\varepsilon}_d}{2}\int_{-\infty}^{\infty}\frac{d\varepsilon}{2\pi}\frac{df(\varepsilon)}{d\varepsilon}A^2(\varepsilon), \tag{34}$$

$$E_d^{(1),WBL}(\varepsilon_d) = -\frac{\hbar\dot{\varepsilon}_d}{2}\int_{-\infty}^{\infty}\frac{d\varepsilon}{2\pi}\varepsilon\frac{df(\varepsilon)}{d\varepsilon}A^2(\varepsilon), \tag{35}$$

$$\dot{N}_d^{(2),WBL}(\varepsilon_d) = -\frac{\hbar\dot{\varepsilon}_d^2}{2}\int_{-\infty}^{\infty}\frac{d\varepsilon}{2\pi}\frac{d^2f(\varepsilon)}{d\varepsilon^2}A^2(\varepsilon), \tag{36}$$

$$\dot{W}^{(2),WBL}(\varepsilon_d) = -\frac{\hbar\dot{\varepsilon}_d^2}{2}\int_{-\infty}^{\infty}\frac{d\varepsilon}{2\pi}\frac{df(\varepsilon)}{d\varepsilon}A^2(\varepsilon), \tag{37}$$

$$\dot{Q}^{(2),WBL}(\varepsilon_d) = -\frac{\hbar\dot{\varepsilon}_d^2}{2}\int_{-\infty}^{\infty}\frac{d\varepsilon}{2\pi}(\varepsilon - \mu)\frac{d^2f(\varepsilon)}{d\varepsilon^2}A^2(\varepsilon), \tag{38}$$

and

$$\dot{S}^{(2)} = \frac{\hbar\dot{\varepsilon}^2}{T}\int_{-\infty}^{\infty}\frac{d\varepsilon}{2\pi}(\varepsilon - \mu)\frac{df(\varepsilon)}{d\varepsilon}\frac{dA^2(\varepsilon)}{d\varepsilon}. \tag{39}$$

These expressions are useful to describe processes involving vanishingly slow (with respect to the typical lead relaxation rate) dot level shifts. Since obtaining higher order analytical expressions rapidly become intractable, for any finite dot driving rate the particle and energy fluxes need to be evaluated from the kinetic equations. In what follows we discuss the numerical evaluation of these functions via the DLvN EOM (Eq. (9)).

*Particle fluxes*

Particle (electron) fluxes flowing between different system segments can be readily evaluated using the DLvN EOM. Given the 1-electron density matrix, $\boldsymbol{\sigma}$, the instantaneous total number of electrons in the entire (dot-lead) system is given by $N_{tot}(t) = \text{Tr}[\boldsymbol{\sigma}(t)]$, which may vary with time due to the electron exchange with the implicit external bath. $N_{tot}$ can be formally divided into contributions from the two system sections as follows: $N_{tot}(t) = N_d(t) + N_L(t)$, where, $N_{d/L}(t) = \text{Tr}[\boldsymbol{\sigma}_{d/L}(t)]$. Since the dot is coupled only to the lead section, the particle flux flowing between the dot and the lead sections can be calculated using Eq. (9) as:

$$J_d(t) = \dot{N}_d(t) = \text{Tr}[\dot{\sigma}_d(t)] = \dot{\sigma}_d(t) = -\frac{i}{\hbar}\sum_l[v_l\sigma_{ld}(t) - \sigma_{dl}(t)v_l] = \frac{2}{\hbar}\sum_l Im[v_l\sigma_{ld}(t)], \quad (40)$$

where the sum runs over all lead state indices, $l$. Similarly, for the lead section we obtain:

$$J_L(t) = \dot{N}_L(t) = \text{Tr}[\dot{\boldsymbol{\sigma}}_L(t)] = -\frac{2}{\hbar}\sum_l Im(v_l\sigma_{ld}(t)) - \Gamma\sum_l[\sigma_{ll}(t) - f(\varepsilon_l)]. \quad (41)$$

As in Eq. (40), the first term on the right hand side of Eq. (41) is the particle flux flowing between the lead and the dot sections, with the opposite sign resulting from the directionality of the current. The second term can be identified as the outgoing particle flux from the lead to the implicit bath, to which it is coupled. Therefore, the particle influx into the bath is given by:

$$J_B(t) = \Gamma\sum_l[\sigma_{ll}(t) - f(\varepsilon_l)]. \quad (42)$$

Eqs. (40)-(42) can also be used to identify the energy resolved particle fluxes [38] that were used in Ref. [4] to evaluate heat currents flowing between the dot and the lead. However, due to technical considerations discussed in Ref. [38] we do not follow this route in the present paper.

*Energy and energy fluxes*

Next, consider the energy fluxes. We first consider the expectation value of the total electronic energy of the entire (dot-lead) system, and express it as a sum of the dot ($E_d$) and lead ($E_L$) components:

$$E_{sys} = tr(\boldsymbol{\sigma H}) = \frac{1}{2}tr(\boldsymbol{\sigma H + H\sigma}) = \frac{1}{2}tr_d(\boldsymbol{\sigma H + H\sigma}) + \frac{1}{2}tr_L(\boldsymbol{\sigma H + H\sigma}) = E_d + E_L, \quad (43)$$

where we have used the cyclic property of the full trace operator and defined

$$\begin{cases} E_d \equiv \frac{1}{2} tr_d(\sigma H + H\sigma) = \frac{1}{2} \langle d|\{\sigma, H\}|d\rangle \\ E_L \equiv \frac{1}{2} tr_L(\sigma H + H\sigma) = \frac{1}{2} \sum_l \langle l|\{\sigma, H\}|l\rangle \end{cases} \quad (44)$$

Note that, since partial traces are used in the definitions of $E_d$ and $E_L$, symmetrization is being employed to assure that the obtained values are real. We also note in passing that the contributions to these expressions from the coupling term of Eq. (4), $V$, are equal (see Appendix B) such that $\frac{1}{2}\langle d|\{\sigma, V\}|d\rangle = \frac{1}{2}\sum_l \langle l|\{\sigma, V\}|l\rangle$. This demonstrates that the contribution of the coupling to the total energy is split evenly between the dot and lead subsystems as was assumed in several recent papers [4,10].

The time variation of the dot's section contribution to the total electronic energy can now be evaluated as:

$$\dot{E}_d = \frac{1}{2}\langle d|\{\dot{\sigma}, H\}|d\rangle + \frac{1}{2}\langle d|\{\sigma, \dot{H}\}|d\rangle \quad (45)$$

The first term on the right-hand-side of Eq. (45) represents energy variations due to the flow of particles in- and out-of the dot, as obtained by the time-derivative of the single-particle density matrix. The second term, which depends on the time-derivative of the Hamiltonian operator, corresponds in the present model to energy variations due to the external time-dependent perturbation acting on the dot section and is related to the work performed on the system, as discussed below. The former can thus be identified with the energy flux entering (or leaving if negative) the dot:

$$J_d^{(E)} = \frac{1}{2}\langle d|\{\dot{\sigma}, H\}|d\rangle = -\frac{i}{2\hbar}\langle d|\{[H,\sigma], H\}|d\rangle - \frac{\Gamma}{4}\langle d|\{\sigma, V\}|d\rangle, \quad (46)$$

where we have used Eq. (9) for $\dot{\sigma}$ and defined (see Eq. (4)):

$$\mathbf{V} \equiv \begin{pmatrix} 0 & v \\ v^\dagger & 0 \end{pmatrix}. \quad (47)$$

Similarly, the energy flux entering the lead section is given by:

$$J_L^{(E)} = \frac{1}{2}\sum_l \langle l|\{\dot{\sigma}, H\}|l\rangle = -\frac{i}{2\hbar}\sum_l \langle l|\{[H,\sigma], H\}|l\rangle - \frac{\Gamma}{4}\sum_l \langle l|\{\sigma, V\}|l\rangle - \Gamma \sum_l \varepsilon_l(\sigma_{ll} - \sigma_{ll}^{eq}), \quad (48)$$

where we have used the fact that $\boldsymbol{\varepsilon}_L$ and $\boldsymbol{\sigma}_L^{eq}$ (whose elements are given by Eq. (10)) are diagonal matrices. We may now define:

$$J_{dL}^{(E)} \equiv -\frac{i}{2\hbar}\langle d|\{[H,\sigma], H\}|d\rangle = \frac{i}{2\hbar}\sum_l \langle l|\{[H,\sigma], H\}|l\rangle = -\frac{i}{2}\sum_l(\varepsilon_d(t) + \varepsilon_l)(v_l\sigma_{ld} - \sigma_{dl}v_l) \quad (49)$$

$$V_d \equiv \langle d|\{\sigma, V\}|d\rangle = \sum_l \langle l|\{\sigma, V\}|l\rangle \quad (50)$$

$$E_{LB} \equiv \sum_l \varepsilon_l (\sigma_{ll} - \sigma_{ll}^{eq}) \tag{51}$$

where the second equality in Eq. (49) stems from the cyclic property the full trace operation: $\langle d|\{[\boldsymbol{H},\boldsymbol{\sigma}],\boldsymbol{H}\}|d\rangle + \sum_l \langle l|\{[\boldsymbol{H},\boldsymbol{\sigma}],\boldsymbol{H}\}|l\rangle = \text{Tr}(\{[\boldsymbol{H},\boldsymbol{\sigma}],\boldsymbol{H}\}) = \text{Tr}([\boldsymbol{H}^2,\boldsymbol{\sigma}]) = 0$, the third equality in Eq. (49) can be obtained from the definition of $\boldsymbol{H}$ (Eq. (7)) and $\boldsymbol{\sigma}$ (Eq. (6)) assuming that $\boldsymbol{v}$ is a real coupling vector, and the last equality in the definition of $V_d$ results from the structure of $\boldsymbol{\sigma}$ and $\boldsymbol{V}$ (Eq. (47)). With these definitions we can rewrite the energy fluxes as:

$$J_d^{(E)} = J_{dL}^{(E)} - \frac{\Gamma}{4} V_d \tag{52}$$

and

$$J_L^{(E)} = -J_{dL}^{(E)} - \frac{\Gamma}{4} V_d - \Gamma E_{LB}. \tag{53}$$

Since the lead section is coupled to both the dot (explicitly) and the bath (implicitly), energy conservation requires that the total energy flux into the lead equals the sum of energy fluxes out of the bath $\left(-J_B^{(E)}\right)$ and the dot sections: $J_L^{(E)} = -J_d^{(E)} - J_B^{(E)}$. Therefore, the energy flux into the implicit bath is given by:

$$J_B^{(E)} = -J_d^{(E)} - J_L^{(E)} = \frac{\Gamma}{2} V_d + \Gamma E_{LB}. \tag{54}$$

The structure of Eqs. (52)-(53) implies that $J_{dL}^{(E)}$ can be identified as the direct energy flux between the dot and the lead sections (from the dot to the lead, when negative). In addition, in the local $(d,\{l\})$ basis employed here, the last term in Eq. (52) and the last two terms in Eq. (53) may be viewed as energy transfer fluxes from the dot and the lead sections into the bath, respectively. Together they account for the overall energy flux into the bath given by Eq. (54). We note that identifying the last term in Eq. (52) as energy flux between the dot and the external bath, which are not directly coupled, is a matter of choice. We could alternatively assign this flux to an additional energy exchange between the dot and the lead, and add and subtract the same term to the RHS of Eq. (53), where they may be thought of as additional fluxes experienced by the lead from the dot and into the bath. Here, it should be emphasized that, regardless of the assignment of the $\frac{\Gamma}{4} V_d$ term to a given system section, there is no time-delay between the associated flux leaving (if positive) the dot and the similar flux entering the external bath.

*Heat fluxes*

Turning to the consideration of heat currents, we encounter the usual conceptual problem associated with the fact that the distinction between energy and heat fluxes involves some form of relaxation. In the quasistatic limit discussed above, no ambiguity arises since this relaxation is assumed to be instantaneous on the driving timescale. When the driving is done at a finite rate, however, the definition of the heat flux becomes ambiguous unless some physical choices are made, as explained below.

Formally, the evolution of the system's (dot-lead) energy, $E_{sys} = \text{Tr}(\boldsymbol{\sigma}(t)\boldsymbol{H}(t))$, when the system undergoes a non-equilibrium process may be cast in the form:

$$\dot{E}_{sys} = \text{Tr}\left(\boldsymbol{\sigma}(t)\dot{\boldsymbol{H}}(t)\right) + \text{Tr}\left(\dot{\boldsymbol{\sigma}}(t)\boldsymbol{H}(t)\right). \tag{55}$$

The first term on the right can be identified as the power (work per unit time) done on the system by the agent that makes the Hamiltonian time-dependent:

$$\dot{W} = \text{Tr}\left(\boldsymbol{\sigma}(t)\dot{\boldsymbol{H}}(t)\right). \tag{56}$$

In our model $\dot{\boldsymbol{H}}(t) = \dot{\varepsilon}_d(t)|d\rangle\langle d|$, so that

$$\dot{W} = \dot{\varepsilon}_d(t)\sigma_{dd}(t) = \dot{\varepsilon}_d(t)n_d(t). \tag{57}$$

Energy conservation implies that the second term in (55) represents the contribution of all other sources of energy change. No such sources exist for a closed system, which in our dot-lead model corresponds to $\Gamma = 0$. Indeed, $\text{Tr}(\dot{\boldsymbol{\sigma}}(t)\boldsymbol{H}(t)) = -i\text{Tr}([\boldsymbol{H}, \boldsymbol{\sigma}]\boldsymbol{H}) = 0$ in such case. Similarly, in a closed system $\dot{N} = \text{Tr}(\dot{\boldsymbol{\sigma}}) = -i\text{Tr}([\boldsymbol{H}, \boldsymbol{\sigma}]) = 0$ expresses conservation of mass.

For an open system ($\Gamma \neq 0$), the second term on the RHS of (55) is determined by the processes that take place at the interface between the lead and the external bath, and expresses the effect of energy and particles exchange between the dot-lead system and this bath. In such a grand-canonical ensemble the total energy variation of the dot-lead system satisfies $\dot{E}_{sys} = \dot{Q} + \dot{W} + \mu\dot{N}$, where we recall that $\mu$ is the chemical potential of the bath. Using Eqs. (7), (9) and (55)-(56) we may write:

$$\dot{Q} + \mu\dot{N} = \dot{E}_{sys} - \dot{W} = \text{Tr}\left[-\Gamma\begin{pmatrix} 0 & \frac{1}{2}\boldsymbol{\sigma}_{d,L} \\ \frac{1}{2}\boldsymbol{\sigma}_{L,d} & \boldsymbol{\sigma}_L - \boldsymbol{\sigma}_L^{eq} \end{pmatrix}\begin{pmatrix} \varepsilon_d(t) & \boldsymbol{v} \\ \boldsymbol{v}^\dagger & \boldsymbol{\varepsilon}_L \end{pmatrix}\right], \tag{58}$$

or

$$\dot{Q} = \text{Tr}\left\{-\Gamma \begin{pmatrix} 0 & \frac{1}{2}\sigma_{d,L} \\ \frac{1}{2}\sigma_{L,d} & \sigma_L - \sigma_L^{eq} \end{pmatrix} \left[\begin{pmatrix} \varepsilon_d(t) & \boldsymbol{v} \\ \boldsymbol{v}^\dagger & \boldsymbol{\varepsilon}_L \end{pmatrix} - \mu \boldsymbol{I}\right]\right\}, \tag{59}$$

where $\boldsymbol{I}$ is the unit matrix in the single electron d-L space and we have used again the fact that the trace of the commutator vanishes.

While formally exact, expressions (58)-(59) may miss the essence of the intended calculation. Since $\boldsymbol{\Gamma}$ was introduced as a mathematical tool that depends on choices made for the finite lead size and the corresponding density of lead states, the heat flux calculated from equation (59) will depend on these unphysical choices. Specifically, we note that in the limit where the lead becomes infinite and its density of states becomes continuous, the dynamics of interest can be described by taking the limit $\boldsymbol{\Gamma} \to \boldsymbol{0}$. In this limit, the calculated heat flux into the implicit bath vanishes (see Eq. (59)), while the lead itself is assumed to remain in thermal equilibrium. Indeed, this is the way the analytical calculation is done, by imposing a thermally equilibrated wide-band lead of uniform density-of-states and a coupling between the dot state and the eigenstates of the free lead, while disregarding any external bath that may be required to affect the lead thermalization. While such a picture of a constantly equilibrated lead may be valid when the dot level shift rate is not too fast, it cannot be implemented for the calculation of heat and particle exchange at finite shift rates within our finite-size model that addresses the lead dynamics explicitly.

An alternative approach for evaluating the heat flux can be devised based on the following two assumptions: (a) The process of interest, namely relaxation by transport of energy and particles between the dot and the lead, is irreversible; and (b) any change in the state of the lead resulting from this transport is insignificant in the sense that it does not affect the relaxation dynamics at the dot-lead interface. The former can be satisfied by taking $\boldsymbol{\Gamma} > \boldsymbol{\rho}^{-1}$, whereas the latter is obtained by choosing a sufficiently large lead model. Under these conditions, the actual relaxation of the lead to its thermal equilibrium by exchange of particles and energy with the external (implicit) bath can occur on a timescale much slower than the timescale of interest, namely the characteristic relaxation time at the dot-lead interface, without affecting the dynamics of interest. In these circumstances, all energy and particle fluxes at the dot-lead interface, being irreversible, are *eventually* expressed as heat and particle currents into the external bath. It is therefore convenient to define a book-keeping procedure associating $\dot{\boldsymbol{N}}$ and $\dot{\boldsymbol{Q}}$ with the fluxes exchanged between dot and the lead (from lead to dot when positive), although their actual realization as fluxes in the external bath might take place on a different, possibly much longer, timescale. Within this picture, the heat current (negative when out of the dot) is defined as:

$$\dot{Q} = J_d^{(E)} - \mu J_d = J_{dL}^{(E)} - \frac{\Gamma}{4} V_d - \mu J_d, \tag{60}$$

where $V_d$ is given by Eq. (50) and $J_d$, $J_d^{(E)}$ and $J_{dL}^{(E)}$ are defined in Eqs. (40), (52), and (49), respectively. Notably, Eq. (60) differs from Eq. (59) by the fact that it considers instantaneous particle and energy fluxes at the dot-lead interface without explicitly involving the bath, while at the same time takes into account the fact that particles entering the lead will be eventually equilibrated to the external chemical potential $\mu$. The difference between the energy carried by the particles as they transfer between the dot and the lead $\left(J_d^{(E)}\right)$ and their energy at equilibrium $(\mu J_d)$ per unit time is then defined as the rate of heat generation in the bath.

Both Eqs. (59) and (60) will be used below. It should, however, be pointed out that neither corresponds exactly to what is calculated in the analytical WBL model [4,7,10,42]. Eq. (59) represents the heat flux exchanged between the system (dot-lead) and an external bath at thermal equilibrium rather than between the dot and the equilibrated lead as in the analytical treatment. Hence, the resulting value of the heat flux depends on the value of $\Gamma$ and will give a physically meaningful result only when $\Gamma$ is taken to represent the correct exchange rate of outgoing and incoming thermal free electrons [34,44]. Eq. (60), on the other hand, is a meaningful representation of a quantity that would eventually be expressed as heat exchanged with the external thermal bath, however, it does not represent the actual heat exchanged in real-time. In this regard, it should be kept in mind that augmenting the finite dot-lead system by coupling it to an external thermal bath was primarily done in order to impose irreversibility on the numerical procedure. It has no other dynamical consequences provided that the lead is taken large enough, namely fluxes at the dot-lead interface should be identical to what they would be if this lead was infinite once the calculation is converged with regards to lead size. Still, we find that the external bath is required for the conceptual definition of the heat flux, although, once the bookkeeping argument is adopted, the heat flux itself is calculated independently of this bath.

Another pertinent numerical point should be mentioned. In what follows we compare analytic results obtained at the WBL with numerical results calculated with a finite-band model. The latter are valid at any driving rate, $\dot{\varepsilon}_d$, whereas the former were obtained only for small $\dot{\varepsilon}_d$ with respect to the relaxation processes that thermalize the lead. While the numerical procedure described above is sufficiently accurate for many applications [29–32,35], the small shift rate corrections to the quasi-equilibrium behavior can be of the order of the small difference between the finite- and infinite-band results. Furthermore, these differences between the finite- and infinite-band results converge slowly with the lead's bandwidth, in particular for the energy calculation (see Eq. (14)). We have discussed these numerical issues in details in Ref. [39], and have found that the following modification to the straightforward numerical calculation provides good agreement with the WBL analytical results. Let

$$\Delta F^{(0)}(\varepsilon_d) \equiv F^{(0)}_{WBL}(\varepsilon_d) - F^{(0)}_{num}(\varepsilon_d) \tag{61}$$

$$\Delta \dot{F}^{(1)}(\varepsilon_d) \equiv \dot{F}^{(1)}_{WBL}(\varepsilon_d) - \dot{F}^{(1)}_{num}(\varepsilon_d) = \dot{\varepsilon}_d \frac{d\Delta F^{(0)}(\varepsilon_d)}{d\varepsilon_d} \tag{62}$$

be, respectively, the differences between the equilibrium values and between the quasi-static change rates of a system property $F = N, E$, calculated analytically in the wide-band limit from Eqs. (13) and (14) and numerically for the corresponding finite bandwidth model. Note that the latter cannot be simply represented by truncating the integrals in Eqs. (13) and (14), because the spectral function $A(\varepsilon)$ of the finite-band model is not simply a truncated version of the infinite-band expression of Eq. (5). We assume that *for any driving speed $\dot{\varepsilon}_d$*, $\Delta \dot{F} = \dot{F}_{WBL} - \dot{F}_{num}$ is well represented by its quasi-static value of Eq. (62). This implies that a numerical calculation aimed to evaluate the effect of finite driving at the WBL can be obtained from:

$$\left[\dot{F}(\dot{\varepsilon}_d, \varepsilon_d) - \dot{F}^{(1)}(\varepsilon_d)\right]_{WBL} \simeq \dot{F}_{num}(\dot{\varepsilon}_d, \varepsilon_d) - \dot{F}^{(1)}_{num}(\varepsilon_d). \tag{63}$$

It should be noted that these flux expressions represent the difference between the quasistatic currents and the instantaneous currents obtained under a finite driving rate. However, care should be exercised when interpreting integrals of these corrections between two values of the externally driven parameter (here $\varepsilon_d$). The reason is that these instantaneous currents do not contain information on the residual relaxation that takes place after the final value of the driving parameter is reached (note that in the sudden limit discussed above this relaxation dominates the non-equilibrium process). At the same time, it is obvious that the integral over the power $\dot{W}$ accounts for the full work associated with the driven process.

*Entropy and entropy fluxes*

Finally, we consider entropy generation and entropy fluxes associated with the dot driving process. Two points should be made at the outset: (i) in the quasistatic limit, the heat and entropy currents are related by $\dot{S}^{(1)}(\varepsilon_d) = \dot{Q}^{(1)}(\varepsilon_d)/T$ and are given in the WBL by Eq. (19); (ii) since the equilibrium entropy is a state-function, any driving protocol that carries the system between two *equilibrium* states corresponding to $\varepsilon_{d1}$ and $\varepsilon_{d2}$ induces the same entropy change $\Delta S^{eq}(\varepsilon_{d1}, \varepsilon_{d2}) = Q^{(0)}(\varepsilon_{d1}, \varepsilon_{d2})/T$, determined by the corresponding quasistatic (reversible) process. Any excess heat generation resulting from the finite driving rate will be eventually expressed in the external bath, after the dot-lead system has reaches its final equilibrium. The total thermodynamic entropy production, expressed as entropy increase in the external bath which is taken to be at equilibrium throughout (i.e. the implicit bath of the numerical model), is:

$$\Delta S(\varepsilon_{d1}, \varepsilon_{d2}|\dot{\varepsilon}_d(t)) = -T^{-1}\big[Q(\varepsilon_{d1}, \varepsilon_{d2}|\dot{\varepsilon}_d(t)) - Q^{(0)}(\varepsilon_{d1}, \varepsilon_{d2})\big]. \tag{64}$$

Here, $Q(\varepsilon_{d1}, \varepsilon_{d2}|\dot{\varepsilon}_d(t))$ is the heat produced under a given dot level shift protocol, $\dot{\varepsilon}_d(t)$, that drives the dot between $\varepsilon_{d1}$ and $\varepsilon_{d2}$, (e.g. $\dot{\varepsilon}_d(t) = (\varepsilon_{d2} - \varepsilon_{d1})\delta(t)$ for a sudden jump at $t = 0$), and the negative sign indicates that this heat was transferred to the external bath. The first law implies that this excess heat is determined by the excess work namely (setting $\mu = 0$):

$$Q(\varepsilon_{d1}, \varepsilon_{d2}|\dot{\varepsilon}_d(t)) - Q^{(0)}(\varepsilon_{d1}, \varepsilon_{d2}) = -W_d(\varepsilon_{d1}, \varepsilon_{d2}|\dot{\varepsilon}_d(t)) - W_d^{(0)}(\varepsilon_{d1}, \varepsilon_{d2}). \tag{65}$$

In departure from Eqs. (64) and (65) that explore thermodynamic changes associated with driving protocols that move a system between two equilibrium states, several recent studies, e.g. [4,7,10,44] have considered the *instantaneous* rates of variations in the particle number, work, heat, as well as entropy, to lowest non-trivial order in the dot driving speed. While such quantities can be calculated, (Eqs. (36)-(38)), their interpretation needs a careful examination as demonstrated by our discussion of the heat flux above. For the model under discussion there is no ambiguity concerning the power (work done or produced per unit-time) and the energy of the dot-lead system as well as its time-derivative are also well defined. With regards to the heat, we saw that we need to distinguished between the rate of heat escaping to the external (implicit) bath in real-time, Eq. (59) and that part of the energy current between the dot and the lead that will eventually be realized as heat in the external bath, Eq. (60).

We can use the bookkeeping argument discussed above for the entropy as well by identifying the excess entropy production as $Q/T$, $Q$ being the excess heat transferred to the external bath calculated from equation (60). Alternatively, we can follow Refs. [4,7,10,42] in calculating the effect of driving on the time-dependent entropy of the dot-lead system defined as:

$$S_{dL}(t) = -k_B \text{tr}\{\boldsymbol{\rho}(t)\ln[\boldsymbol{\rho}(t)]\}, \tag{66}$$

where $\boldsymbol{\rho}$ is the many electron density matrix. The following interesting result was obtained for the lowest order excess quantity [4,7,10,44]:

$$\dot{S}_{dL}^{(2)} = \frac{1}{T}\left(\dot{W}_d^{(2)} + \dot{Q}^{(2)}\right), \tag{67}$$

which serves to identify $\frac{\dot{W}_d^{(2)}}{T}$ as the rate of entropy production. Indeed, Eq. (67) expresses the fact that this entropy production (positive if work is done on the system) is expressed as a sum of a term $-\frac{\dot{Q}^{(2)}}{T}$ (positive when heat exits the system) associated with the excess heat given to the external bath plus the excess change in the system (dot-lead) entropy, $\dot{S}_{dL}^{(2)}$. With this interpretation $\dot{Q}^{(2)}$ has to be the heat calculated from Eq. (59), which, as explained above, is model dependent and not a physically meaningful quantity. Furthermore, as was shown in Ref. [15], the result (67) cannot be extended to higher orders in the driving speed, so our finite rate calculation is not expected to rigorously show it.

In what follows, we examine our ability to numerically evaluate quantities analogous to those appearing in Eq. (67), namely to use the time-dependent single-electron density matrix obtained from the DLvN EOM to evaluate the deviations of the work, heat, and system (dot-lead) entropy rates from their quasistatic values obtained from Eq. (62). Below, these deviations are referred to as the excess (with respect to the reversible process) work, heat, and entropy rates. The excess power is straightforward to obtain using Eq. (18) in the form:

$$\dot{W}_d^{excess}(t) = \dot{\varepsilon}_d(t)\left[\sigma_d(t) - \sigma_d^{(eq)}(\varepsilon_d(t))\right], \tag{68}$$

$\sigma_d^{(eq)}(\varepsilon_d(t))$ being the equilibrium dot occupation at a given dot position. For the heat flux we need to calculate:

$$\dot{Q}^{excess}(t) = \dot{Q}(\dot{\varepsilon}_d, \varepsilon_d) - \dot{\varepsilon}_d \lim_{\dot{\varepsilon}_d \to 0} \dot{\varepsilon}_d^{-1}\dot{Q}(\dot{\varepsilon}_d, \varepsilon_d). \tag{69}$$

Note that in Eq. (69) $\lim_{\dot{\varepsilon}_d \to 0} \dot{\varepsilon}_d^{-1}\dot{Q}(\dot{\varepsilon}_d, \varepsilon_d)$ is the first derivative of the heat flux with respect to the driving rate that, when multiplied by $\dot{\varepsilon}_d$, provides the quasistatic contribution to the heat flux.

To calculate the corresponding entropy of the dot-lead system we evaluate Eq. (66) for the many-body density matrix $\boldsymbol{\rho}$, which needs to be expressed it in terms of the single-electron density matrix $\boldsymbol{\sigma}(t)$ of Eq. (9). In Appendix A we show that the resulting expression for non-interacting systems is given by:

$$S(t) = -k_B \sum_j \{s_j(t)\ln[s_j(t)] + [1 - s_j(t)]\ln[1 - s_j(t)]\}, \tag{70}$$

where $\{s_j(t)\}$ is the set of eigenvalues of $\boldsymbol{\sigma}(t)$. At equilibrium $s_j \to f(\varepsilon_j(\varepsilon_d); \beta, \mu)$, where $\{\varepsilon_j(\varepsilon_d)\}$ are the eigenvalues of the Hamiltonian of the dot-lead system at a given $\varepsilon_d$. In this limit Eq. (70) reduces to its equilibrium counterpart. Under a finite dot-level shift rate the excess entropy is given by:

$$S^{\text{excess}}(t) = S(t) - S^{eq}(\varepsilon_d(t)), \tag{71}$$

where $S^{eq}(\varepsilon_d(t))$ is calculated via Eq. (70) by diagonalizing the dot-lead Hamiltonian for any given value of $\varepsilon_d$.

To conclude this section, we note that other entropy expressions, associated with the inner (dot) system only, can be considered. One can adopt for the dot entropy $S_d$ the definition of Esposito et al. [45]:

$$S_d(t) = -k_B \text{Tr}_d[\rho_d \ln \rho_d], \tag{72}$$

where $\rho_d = \text{Tr}_B \rho$ and B is the total system without the dot. However, properties derived for the evolution of this function assume an uncorrelated initial state, $\rho(0) = \rho_d(0)\rho_{B,eq}$ [45]. Alternatively, one can follow Ingarden [46], in focusing on the quantum mechanical operator entropy, $S_Q$, associated with a Hermitian operator $\hat{Q}$. $S_Q$ quantifies the amount of missing information about the observable $Q$ in a given system state $\psi$. Following Ben Naim [47] (see also Lent [48]), for such an operator written in its eigenbasis representation, $\hat{Q} = \sum_q q |\phi_q\rangle\langle\phi_q|$, the operator entropy is defined in terms of the probability $P_q = |\langle\phi_q|\psi\rangle|^2$ as $S_q = -\sum_q P_q \log_2 P_q$. For the operator $\hat{Q} = \hat{c}_d^\dagger \hat{c}_d$, the corresponding operator entropy is the binary entropy associated with the dot,

$$S_d(t) = -p_d(t) \log_2[p_d(t)] - [1 - p_d(t)] \log_2[1 - p_d(t)] \tag{73}$$

where $p_d = \sigma_{dd}$ is the probability that the dot is occupied.

## 4. Results

**The Quasi-Static and Sudden Jump Limits**

In this section, we present and discuss numerical results calculated for the quasistatic, sudden jump, and finite driving rate non-equilibrium processes. These results were obtained from the model of Eqs. (2)-(4) and Figure 1, where the dot energy, $\varepsilon_d$, is shifted at a constant rate $\dot{\varepsilon}_d$. Unless otherwise specified, the following set of parameters were used: energy is expressed in units of $\hbar\gamma$ so that $\hbar\gamma = 1$, $\hbar\Gamma = \Delta\varepsilon = 0.1$, $k_B T = 0.5$, $BW$ (bandwidth) $= 10$ (setting the number of lead states to be $N_L = \frac{BW}{\Delta\varepsilon} = 100$). In what follows, we also take $\hbar = 1$, hence $\gamma = 1$, the time unit is $\gamma^{-1} = 1$, and $\dot{\varepsilon}_d$ is given in units of $\hbar\gamma^2 = 1$.

We start by considering the extreme limits of quasi-static driving and a sudden jump. As discussed above, the calculation of these equilibrium functions can be done by diagonalizing the Hamiltonian and occupying the eigen-levels according to the Fermi-Dirac distribution, or by calculating the 1-electron equilibrium density matrix for the dot-lead system by solving Eq. (9) for $\dot{\sigma} = 0$, which may be recast in a Sylvester-type equation [25]. Results of both procedures are shown in red and blue lines, respectively, in Figures 2-4 below. Also shown are results of the analytical wide band expressions (dashed-black lines) [4]. Fig. 2 shows the dot occupation $n_d$ (panel a), the $\varepsilon_d$-dependent contribution to the energy, $E_d$ (panel b), and the entropy $S$ (panel c), all displayed as deviations from their values at a reference equilibrium state, here taken at $\varepsilon_{d1} = -3$, plotted against $\varepsilon_d$. Being state functions, the values displayed do not depend on the process that leads from $\varepsilon_{d1}$ to $\varepsilon_d$. Figs. 3 and 4 show, for a quasi-static driving and sudden jump, respectively, the work (panels a) and the heat (panels b) plotted as well as functions of $\varepsilon_d$. Note that in the sudden jump case, work and heat are produced in different parts of the dynamic evolution: work during the jump and heat (as well as chemical energy) during the subsequent relaxation.

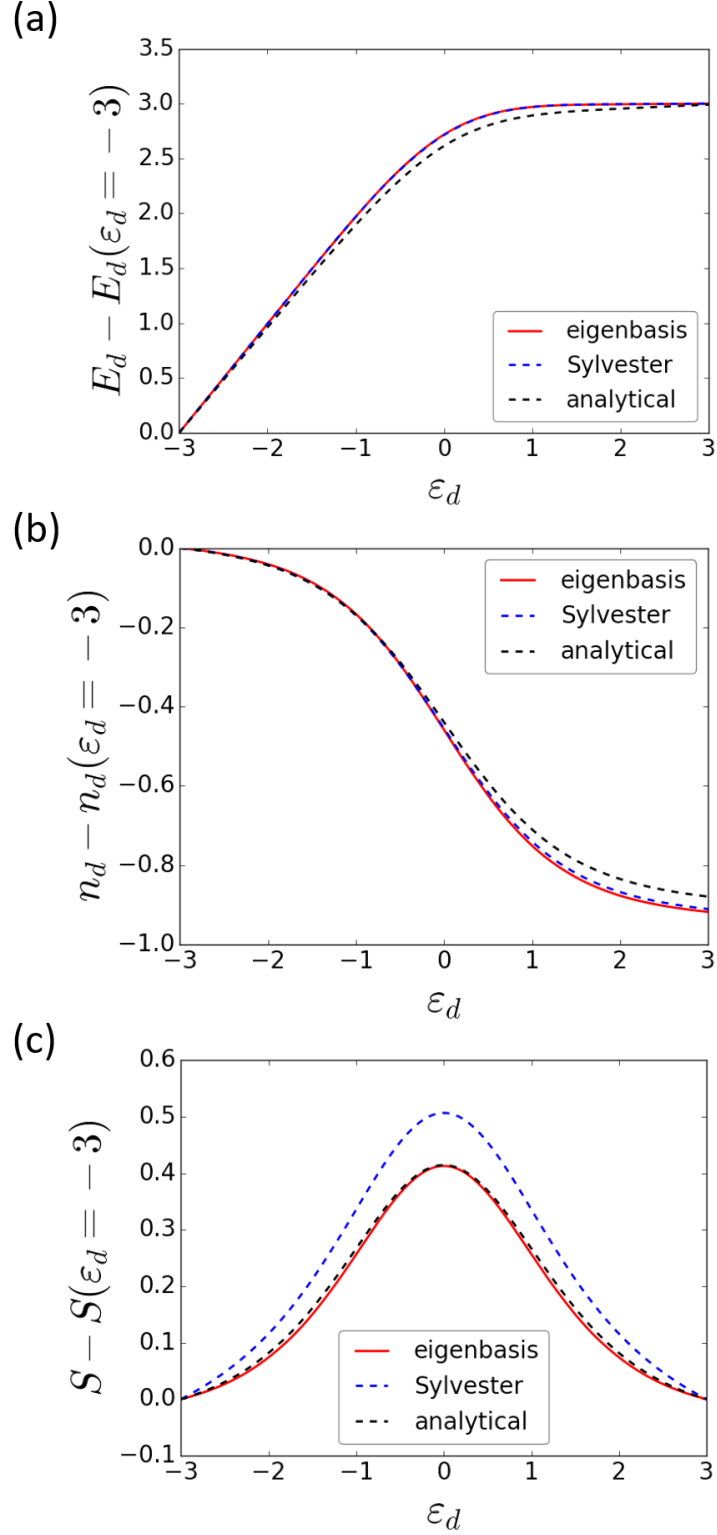

Figure 2: Equilibrium dot occupation (panel (a), calculated using Eq. (23)), energy (panel (b), calculated using Eq. (24)) and entropy (panel (c), calculated using Eq. (22)), measured relative to their values at $\varepsilon_{d1} = -3$, plotted against the dot energy $\varepsilon_d$ as calculated by assigning Fermi-Dirac occupations to the eigenstates of the dot-lead system Hamiltonian (red lines) or based on the 1-electron equilibrium density matrix obtained by solving the Sylvester equation (dashed blue lines) [25]. The analytical WBL expressions (Eqs. (13)-(15)) are represented by the dashed black line.

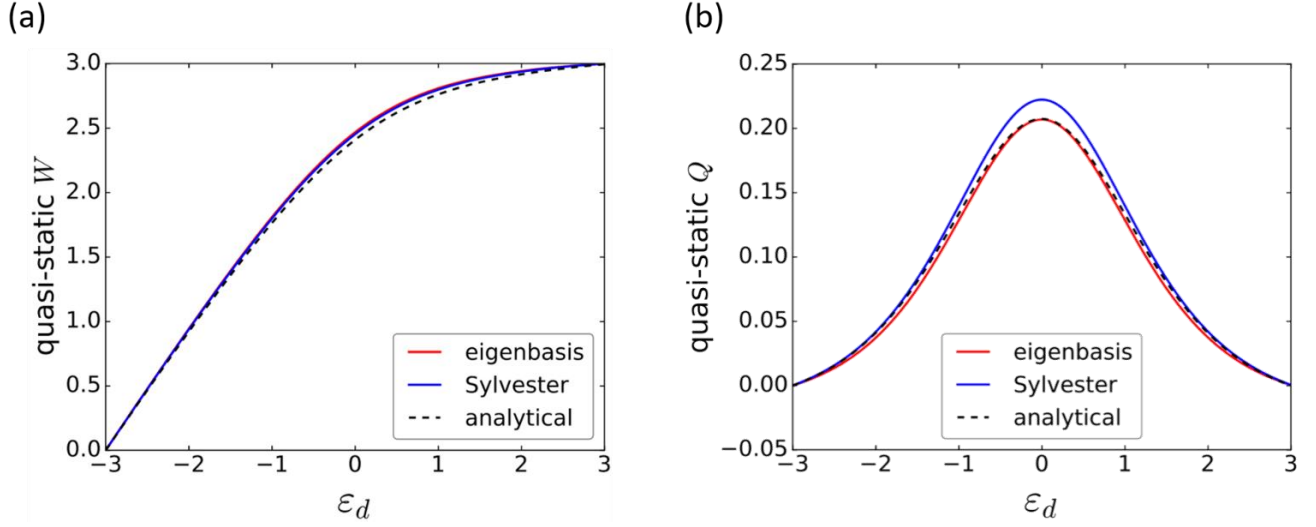

Figure 3: The work (a) calculated using Eq. (26) and the heat (b) calculated from Eq. (27) obtained for quasi-static driving of the dot energy from $\varepsilon_{d1} = -3$ to $\varepsilon_d$. Color scheme is the same as in Fig. 1.

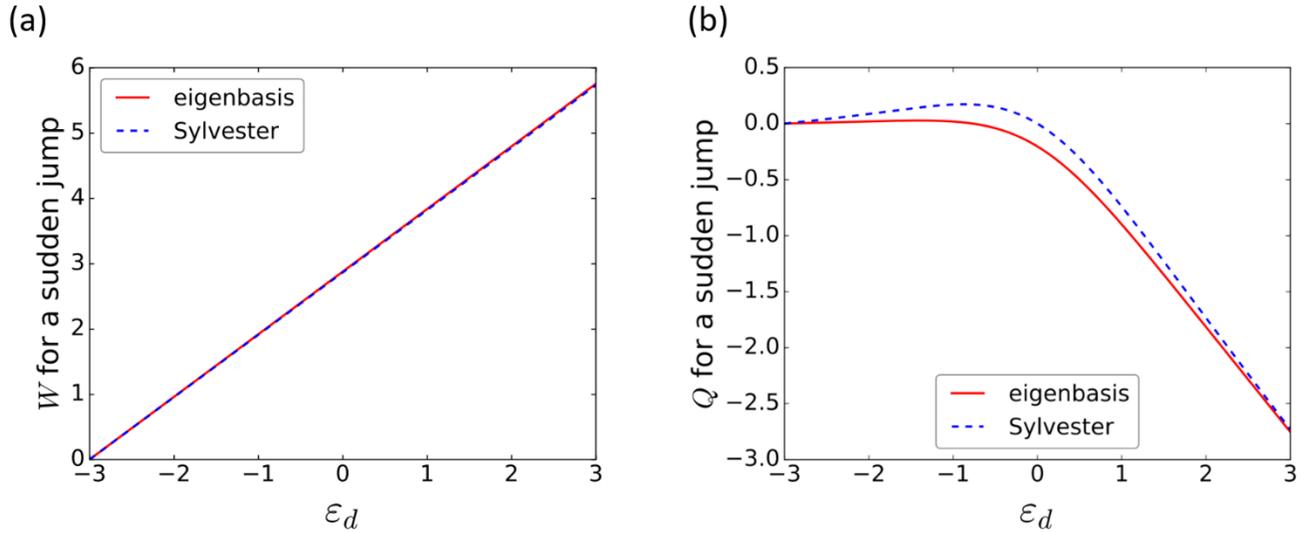

Figure 4: (a) The work (calculated using Eq. (29)) and (b) the heat (calculated using Eq. (30)) obtained for a sudden jump of the dot level from $\varepsilon_{d1} = -3$ to $\varepsilon_d$. Red lines: results obtained by assigning Fermi-Dirac occupations to the eigenstates of the dot-lead system Hamiltonian. Blue lines: Results based on the 1-electron equilibrium density matrix obtained by solving the Sylvester equation [25].

These results generally show good agreement between the different calculations, implying that our finite numerical model provides a reasonable representation of the wide-band limit, and that the Sylvester solution provides a decent approximation to the exact equilibrium density matrix. Note that small deviations do exist and are more appreciable in the heat and entropy differences calculations due to their small numerical values. These discrepancies are expected to become smaller as the size of the numerical basis increases, i.e. when a larger lead model is used (see SI). We note that Figs. 2-4 show changes in thermodynamic functions assigned to the dot subsystem. We could alternatively show the changes in the corresponding quantities associated with the full dot-lead system. For non-interacting electrons and in the wide band limit the results for these changes should be identical. However, small differences are obtained for the present finite system model (see section 4 of the Supporting Information) because the system's spectral function slightly depends in this case on the dot level energy.

**Finite Dot-Level Driving Rate**

Next, consider shifting the dot's energy at a finite rate. Some preliminary notes are in place: (i) As mentioned above, recent analytical studies [4–7,10–15] have evaluated the lowest order (in the driving rate) corrections to thermodynamic functions and fluxes. To approach this limit in the numerical simulations we need to use relatively slow shift rates, for which deviations from the quasistatic limit are small, implying relatively large numerical errors. Such numerical errors become smaller at higher driving rates, where the numerical approach is obviously most useful. Still, to facilitate comparison with analytical results we have chosen to display, for any thermodynamic function $F$, the normalized excess flux:

$$\frac{\dot{F}^{excess}}{\dot{\varepsilon}_d^2} = \frac{\dot{F} - \dot{F}_{eq}}{\dot{\varepsilon}_d^2} \tag{74}$$

(note that for state functions $\dot{F}_{eq}/\dot{\varepsilon}_d = \partial F_{eq}/\partial \varepsilon_d$). (ii) Since in non-equilibrium calculations we use Eq. (9) for the dynamical evolution of the 1-electron density matrix, it makes sense to use the Sylvester equation [25] for the reference equilibrium matrix, expecting errors due to the approximate nature of the equilibrium state enforced via the thermal boundary conditions to at least partially cancel. (iii) In contrast to the results displayed above, which considered changes in thermodynamic functions when the system evolves between two equilibrium states, the results shown below correspond to non-equilibrium thermodynamic fluxes obtained under constant driving rate.

Figures 5-9 show these normalized excess thermodynamic fluxes, $\dot{F}^{excess}/\dot{\varepsilon}_d^2$, with $F = N, E, W, Q$ and $S$, plotted against $\varepsilon_d$ for two finite shift rates, $\dot{\varepsilon}_d$. To obtain these results we have calculated numerical derivatives of the expressions for the thermodynamic functions, Eqs. (28), (44), and (70), respectively, and subtract the numerical derivative of the corresponding Sylvester equation-based equilibrium values. The rates of producing (or absorbing) work and heat are directly given by Eqs. (68) and (69).

Focusing first on the dot's population, energy, and the work performed to move its level, we find that for slow dot-level shift velocities the corresponding excess thermodynamic fluxes show good agreement with the analytical results. Small differences can be attributed to contributions beyond first order and to the difference between the dynamics of our relatively small model system with a discrete lead level spectrum and the analytical resonant level model in the wide band limit. More important are the deviations associated with increasing dot level shift rates, that reflect the fact that relaxation to equilibrium lags behind the evolution of $\varepsilon_d(t)$. This demonstrates the capability of our numerical scheme to explore dynamical regimes that are difficult to access using analytical treatments.

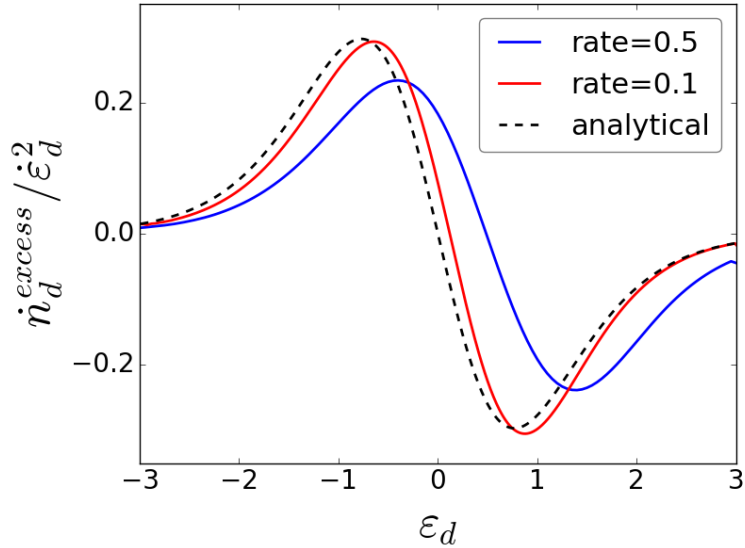

Figure 5: The normalized excess dot occupation variation rate, calculated using Eq. (40), as a function of $\varepsilon_d$ for $\dot{\varepsilon}_d = 0.5$ (full blue line) and $\dot{\varepsilon}_d = 0.1$ (full red line) compared to the analytical solution (Eq. (36); dashed black line).

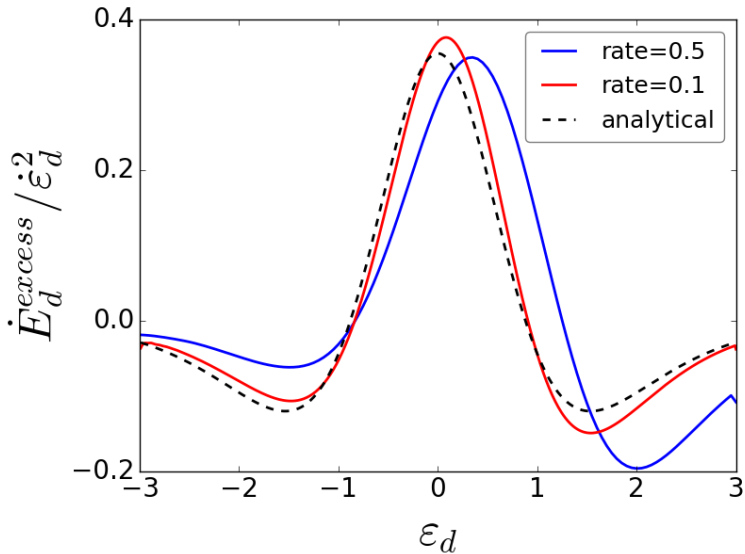

Figure 6: The normalized excess dot energy variation rate, calculated via Eq. (62), as a function of $\varepsilon_d$ for $\dot{\varepsilon}_d = 0.5$ (full blue line) and $\dot{\varepsilon}_d = 0.1$ (full red line) compared to the analytical solution (Eqs. (36), (37), (38); dashed black line).

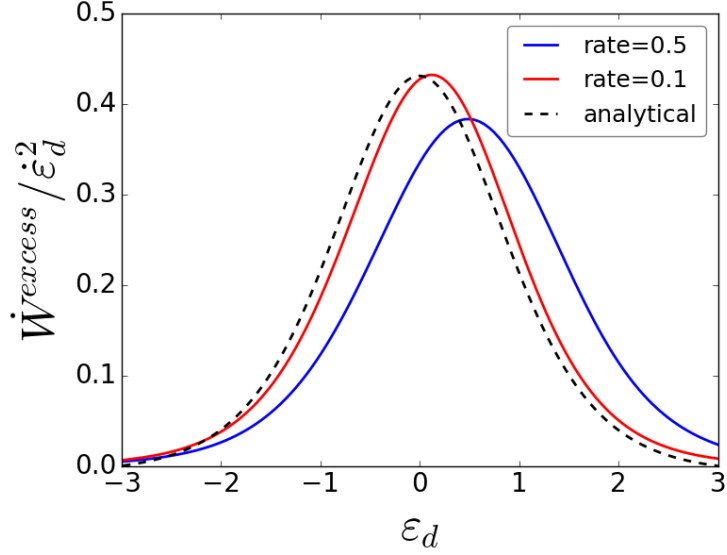

Figure 7: The normalized excess performed work rate, calculated using Eq. (62), as a function of $\varepsilon_d$ for $\dot{\varepsilon}_d = 0.5$ (full blue line) and $\dot{\varepsilon}_d = 0.1$ (full red line) compared to the analytical solution ((37); dashed black line).

For the heat flux, calculated using Eqs. (60) and (69) (Fig. 8a), we also find qualitative agreement with the analytical results at a slow driving rate and a similar lag behavior of the excess heat flux with increasing dot energy shift rate. Here, as well, the deviations between the numerical and analytical results can be associated with higher order contributions in the dynamical results and to the finite discrete nature of the numerical model. Another source of deviation is the above-mentioned approximate nature of the relaxation imposed on the system, driving the zero-order lead levels rather than the exact system eigenstates, to equilibrium. Naturally, the resulting differences between the numerical and analytical results are expressed more strongly in the heat calculation due to its overall smaller absolute magnitude obtained under the studied conditions. As discussed above, the results obtained by using Eq. (59) instead of (60) for the heat flux (Fig. 8b) represent the actual heat exchanged instantaneously between the dot-lead system and the implicit bath in the finite numerical model. These results do not correspond to the heat exchanged between the dot subsystem and the wide-band lead adopted in the analytical treatment and they depend on the choice of the driving rate $\Gamma$, which in turn is related to the discrete spectrum of the specific finite lead model [34].

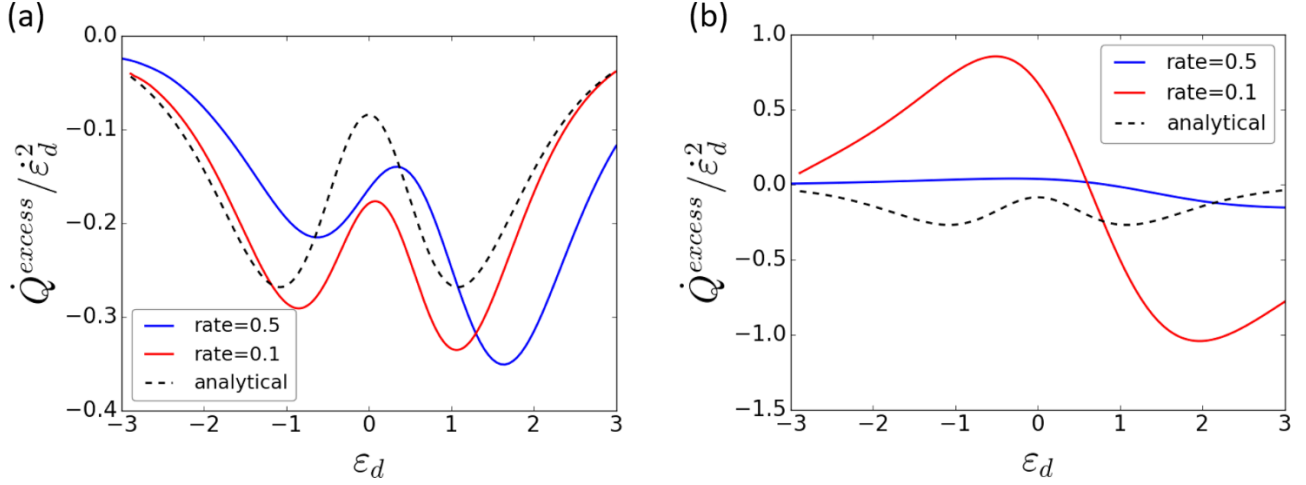

Figure 8: (a) The normalized excess produced heat rate, calculated using Eqs. (60) and (69), plotted as a function of $\varepsilon_d$ for $\dot{\varepsilon}_d = 0.5$ (full blue line) and $\dot{\varepsilon}_d = 0.1$ (full red line), compared to the analytical solution (Eq. (38); dashed black line). (b) Similar results calculated using Eq. (59) and (69) (note the different y-axis scales such that the dashed line representing the analytical WBL result is identical in both figures).

Finally, in Figure 9 we present the excess entropy change rate calculated via Eqs. (70) and (71) as a function of the dot position for the two dot level shift rates considered. As can be seen, even at the lower shift rate considered the agreement between the numerical and analytical results is not satisfactory. While part of the deviation can still be associated with the comparison between analytical wide band approximation results and finite discrete band numerical calculations and with contributions beyond first order in the latter, one important point should be considered. The analytical model focuses on the *dot* contribution to the system's entropy, whereas Eq. (70) evaluates the entropy from the instantaneous occupation of the *dot-lead* system eigenstates. To obtain better agreement between the two results one can project the dot-lead entropy expression of Eq. (70) on the dot state:

$$S_d(t) = -k_B \sum_j |\langle d|j\rangle|^2 \{s_j(t)\ln[s_j(t)] + [1 - s_j(t)]\ln[1 - s_j(t)]\}. \tag{75}$$

The results obtained using Eq. (75) are presented in Fig. 10, showing better agreement between the low driving rate numerical (full red line) and first-order analytical (dashed black line) excess entropy and the expected lag at higher driving rates (full blue line). For comparison purposes, we plot also results obtained using the binary (information) entropy expression associated with the dot state, given by Eq. (73). The obtained results (dashed red and blue lines in Figure 10) are found to be in good agreement with the dot-projected results of Eq. (75). This indicates that the two local entropy expressions of Eqs. (73) and (75) account for most of the $\varepsilon_d$ dependent entropy contribution evaluated by the analytical treatment of Ref. [4].

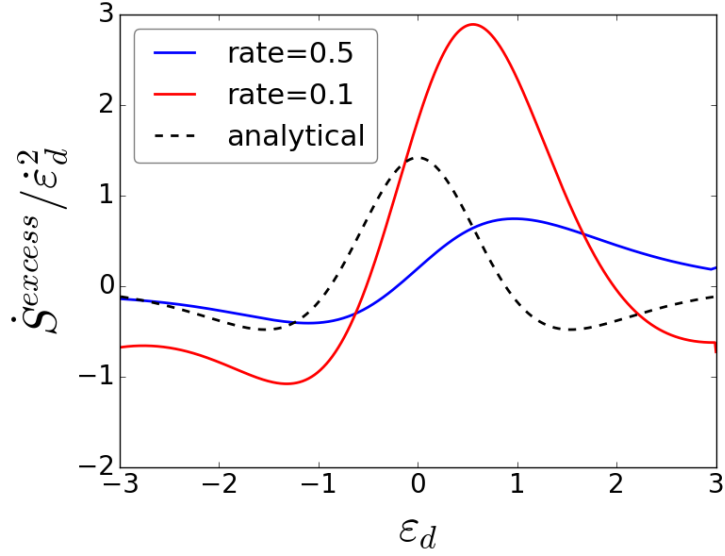

Figure 9: The normalized excess total (dot-lead) entropy change rate, calculated via Eqs. (70) and (71), as a function of $\varepsilon_d$ for $\dot{\varepsilon}_d = 0.5$ (full blue line) and $\dot{\varepsilon}_d = 0.1$ (full red line), compared to the analytical solution (Eq. (39); dashed black line).

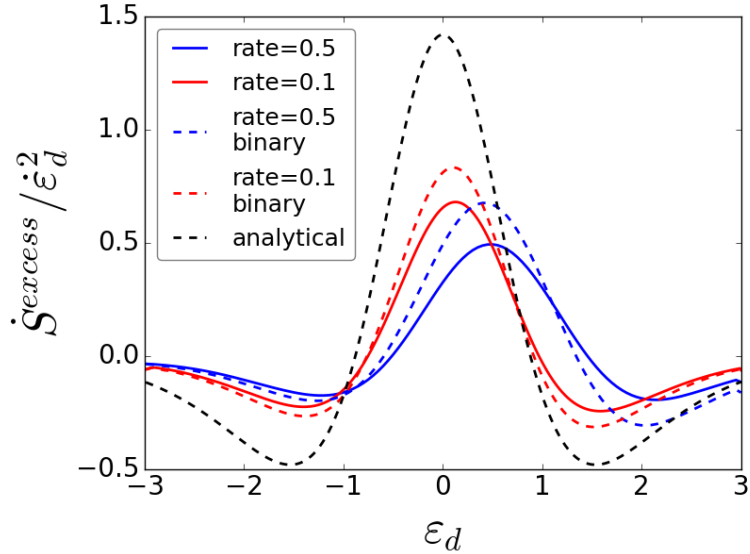

Figure 10: The normalized "dot entropy" change rate plotted against $\varepsilon_d$ for $\dot{\varepsilon}_d = 0.5$ (blue) and $\dot{\varepsilon}_d = 0.1$ (red). Solid lines represent the normalized dot-projected entropy of Eq. (75). Dashed lines correspond to the information (binary) entropy associated with the dot state, calculated via Eq. (73). The analytical WBL expression (Eq. (39)) is displayed by the dashed black line, and is the same as in Fig. 9.

## **Conclusions**

The results presented above demonstrate the suitability of the Driven Liouville von-Neumann methodology for the study of non-equilibrium thermodynamic properties of open quantum systems, even in the regime of strong coupling between the subsystem of interest and its environment. Specifically, we have focused on the resonant level model subjected to a time-dependent driving of the dot energy, but the same numerical approach can obviously be applied to a variety of other and more realistic models. Unlike recent analytical treatments of this problems, which rely on expansions in powers of the driving rate and are therefore limited to slow driving scenarios, our numerical approach applies also to systems subjected to high driving rates and, in fact, becomes more efficient and accurate in this regime. Furthermore, the numerical model is not limited to the wide band limit often used to simplify analytical treatments and can be used with arbitrary lead band structure. Here, however, we have deliberately attempted to stay close to the wide band and slow driving limits, at which the available analytical results are reliable, in order to facilitate comparison.

Our numerical results for the dot occupation, energy, and entropy as well as for the work and heat fluxes show excellent agreement with the analytical theory in the quasi-static limit. Furthermore, the DLvN predictions (with the exception of entropy) correspond well to the analytical results also for the case of finite slow dot level driving - the most challenging regime for our numerical simulations. We therefore conclude that the DLvN approach can be used as a complementary tool to analytical calculations, providing valuable information in dynamic and thermodynamic regimes that are difficult to explore by current analytical treatments.

One important aspect of applying thermodynamics to non-equilibrium situations clearly comes to light in the numerical analysis. In contrast to the energy, particle number, and work, the definitions of heat and entropy in non-equilibrium situations is subject to the usual uncertainty encountered whenever the bath or part of it (here the lead) is simulated explicitly as part of the system. In the present calculation, while the heat exchanged with the external bath and the entropy of the dot-lead system are readily evaluated, their evolution is determined by the DLvN driving rate ($\Gamma$) that is used as a tool to stabilize the numerical solution and depends on the choice of lead model size. Since this choice is somewhat arbitrary and depends on the convergence of the results, the physical meaning of the $\Gamma$ dependent heat and entropy expressions become model dependent. More generally, the numerical calculation sheds light on the ambiguity associated with the fact that particle and energy fluxes exchanged by a system of interest and the realization of these fluxes as thermodynamic variations in an equilibrated external bath occur on different timescales. Alternative local estimates that focus on changes that take place at the dot-lead interface were explored. For example, the reasonable assumption that all the dot outgoing energy eventually translates into heat at the equilibrated bath was used to define, as a bookkeeping tool, an instantaneous heat flux out of the dot. For the entropy, local quantities such as the system entropy projected on the dot and the dot information entropy appear to be useful. It should be emphasized that these aspects of our numerical results reflect not any difficulty in evaluating the time evolution of the driven system but only the ambiguity in interpreting these results in a thermodynamic framework.

Finally, the present numerical work has focused on a non-interacting particles (here fermions) model. Generalization of the numerical procedure to better descriptions of the system, e.g. interacting electrons, will be explored in future studies.

## Acknowledgement

The research of AN is supported by the Israel-U. S. Binational Science Foundation, the German Research Foundation (DFG TH 820/11-1), the U. S. National Science Foundation (Grant No. CHE1665291), and the University of Pennsylvania. OH is grateful for the generous financial support of the Israel Science Foundation under grant number 1740/13 and the Center for Nanoscience and Nanotechnology of Tel-Aviv University. AO gratefully acknowledges the support of the Adams Fellowship Program of the Israel Academy of Sciences and Humanities, and the Naomi Foundation through the Tel-Aviv University GRTF Program.

# Appendix A: Entropy in Terms of the Single-Particle Density Matrix.

In terms of the its many-electron density matrix operator, $\boldsymbol{\rho}$, the entropy of a system is given by:

$$S = -k_B \text{Tr}[\boldsymbol{\rho} \ln(\boldsymbol{\rho})]. \tag{A.1}$$

For a system of non-interacting electrons, Wick's theorem implies that $\boldsymbol{\rho}$ can be written as an exponential of a free Fermion operator [42]:

$$\boldsymbol{\rho} = K e^{-\beta \widehat{A}}, \tag{A.2}$$

where $K$ is a scalar, $\beta = (k_B T)^{-1}$, $k_B$ is Boltzmann's constant, $T$ is the temperature, and the many-electron operator $\widehat{A}$ is given by:

$$\widehat{A} = \sum_{i,j} A_{ij} c_i^\dagger c_j. \tag{A.3}$$

Here, $c_i^\dagger$ and $c_i$ are the single-electron creation and annihilation operators in a general (not necessarily diagonal) representation of the matrix $\boldsymbol{A}$ formed from the coefficients $A_{ij}$, where it should be noted that the matrix $\boldsymbol{A}$ is not the matrix representation of the operator $\widehat{A}$ in the basis of its eigenstates.

Using Eqs. (A.1), (A.2), and (A.3) the entropy can thus be written as:

$$S = -k_B \left[ \ln(K) - \beta \text{Tr}(\boldsymbol{\rho} \widehat{A}) \right], \tag{A.4}$$

where we have used the fact that $tr(\boldsymbol{\rho}) = 1$. In order to write Eq. (A.4) in terms of the single-electron density matrix $\boldsymbol{\sigma}$ one needs to find an expression for the pre-factor $K$ and a relation between the matrices $\boldsymbol{A}$ and $\boldsymbol{\sigma}$.

We start by finding an explicit expression for the pre-factor $K$. To this end, we consider the unitary transformation matrix $\boldsymbol{T}$ that diagonalizes $\boldsymbol{A}$. $\boldsymbol{T}$ is constructed from the column eigenvectors of the matrix $\boldsymbol{A}$, which we denote by $\{\boldsymbol{\phi}_k\}$. Given $\boldsymbol{T}$ we can transform the Fermion operators according to:

$$\tilde{\boldsymbol{c}}^\dagger = \tilde{\boldsymbol{a}}^\dagger \boldsymbol{T} \; ; \boldsymbol{a} = \boldsymbol{T}^{-1} \boldsymbol{c}, \tag{A.5}$$

where $\tilde{\boldsymbol{c}}^\dagger$ denotes a row vector of creation operators, $c_i^\dagger$, given in the general representation of the matrix $\boldsymbol{A}$, $\tilde{\boldsymbol{a}}^\dagger$ denotes a row vector of creation operators, $a_i^\dagger$, given in its eigenbasis representation and $\boldsymbol{c}$ and $\boldsymbol{a}$ are column vectors of the corresponding annihilation operators. The elements of the matrix $\boldsymbol{A}$ itself can be written in terms of its eigenvalues, $\varepsilon_k$, and eigenvectors as follows:

$$A_{ij} = \sum_k \varepsilon_k \phi_{ik} \phi_{jk}^*, \tag{A.6}$$

where $\phi_{ik}$ is the $i^{th}$ element of the column vector $\boldsymbol{\phi}_k$.

We can write Eq. (A.2) in the diagonal representation as follows:

$$\boldsymbol{\rho} = K e^{-\beta \sum_k \varepsilon_k a_k^\dagger a_k} = K \prod_{k=1}^{M} e^{-\beta \varepsilon_k a_k^\dagger a_k}, \qquad (A.7)$$

where $M$ is the number of states and we have used the fact that all the operators in the exponents are commutable to write the corresponding Kronecker product.

The pre-factor $K$ can now be found from the condition $\text{Tr}(\boldsymbol{\rho}) = 1$ by using the fact that the trace of a Kronecker product of two matrices is the multiplication of the traces of the individual matrices and that for spinless Fermionic systems $\text{Tr}\left(e^{-\beta \varepsilon_k a_k^\dagger a_k}\right) = 1 + e^{-\beta \varepsilon_k}$, where the two terms on the right-hand-side stand for the unoccupied and occupied $k^{th}$ eigenstate, respectively. The resulting expression is:

$$K = \prod_{k=1}^{M} \left(1 + e^{-\beta \varepsilon_k}\right)^{-1}. \qquad (A.8)$$

Next, to find a relation between $\boldsymbol{A}$ and $\boldsymbol{\sigma}$ we show that the two matrices share the same eigenbasis. The elements of the single-electron density matrix are given by the expectation value of the correlation function of the Fermionic operators. This relates the single- and many-electron density matrices as follows:

$$\sigma_{ji} = \langle c_i^\dagger c_j \rangle = \text{Tr}(\boldsymbol{\rho} c_i^\dagger c_j). \qquad (A.9)$$

Substituting equations (A.7) and (A.8) in (A.9) yields:

$$\sigma_{ji} = \text{Tr}(\boldsymbol{\rho} c_i^\dagger c_j) = \text{Tr}\left[\left(K e^{-\sum_k \beta \varepsilon_k a_k^\dagger a_k}\right)\left(\sum_n \phi_{ni}^* a_n^\dagger\right)\left(\sum_m \phi_{mj} a_m\right)\right] =$$
$$K \sum_n \sum_m \phi_{ni}^* \phi_{mj} \text{Tr}\left(e^{-\sum_k \beta \varepsilon_k a_k^\dagger a_k} a_n^\dagger a_m\right). \qquad (A.10)$$

Using Eq. (A.8) for $K$ and writing the exponent as a Kroenecker product we get:

$$\sigma_{ji} = \left[\prod_{k=1}^{M}(1 + e^{-\beta \varepsilon_k})^{-1}\right] \sum_n \sum_m \phi_{ni}^* \phi_{mj} \text{Tr}\left[\left(\prod_k^M e^{-\beta \varepsilon_k a_k^\dagger a_k}\right) a_n^\dagger a_m\right] = \left[\prod_{k=1}^{M}(1 + e^{-\beta \varepsilon_k})^{-1}\right] \sum_n \phi_{ni}^* \phi_{nj} \text{Tr}\left[\left(\prod_k^M e^{-\beta \varepsilon_k a_k^\dagger a_k}\right) a_n^\dagger a_n\right], \qquad (A.11)$$

where we have used the fact that the trace over the product $a_n^\dagger a_m$ vanishes unless $n = m$. Taking explicitly the trace over all elements $k \neq n$ gives:

$$\sigma_{ji} = \left[\prod_{k=1}^{M}(1+e^{-\beta\varepsilon_k})^{-1}\right]\sum_n \phi_{ni}^*\phi_{nj}\text{Tr}\left[\left(\prod_{k\neq n}^{M} e^{-\beta\varepsilon_k a_k^\dagger a_k}\right)e^{-\beta\varepsilon_n a_n^\dagger a_n}a_n^\dagger a_n\right] = \left[\prod_{k=1}^{M}(1+e^{-\beta\varepsilon_k})^{-1}\right]\sum_n \phi_{ni}^*\phi_{nj}\left[\prod_{k\neq n}^{M}(1+e^{-\beta\varepsilon_k})\right]\text{Tr}\left[e^{-\beta\varepsilon_n a_n^\dagger a_n}a_n^\dagger a_n\right] \tag{A.12}$$

The remaining trace yields $\text{Tr}\left[e^{-\beta\varepsilon_n a_n^\dagger a_n}a_n^\dagger a_n\right] = 0 + e^{-\beta\varepsilon_n}$, where the null and the exponent on the right-hand-side stand for the unoccupied and occupied $n$ state, respectively. The resulting expression for the elements of the single-electron density matrix therefore is:

$$\sigma_{ji} = \sum_n \phi_{ni}^*\phi_{nj}\frac{e^{-\beta\varepsilon_n}}{1+e^{-\beta\varepsilon_n}} = \sum_n \frac{1}{e^{\beta\varepsilon_n}+1}\phi_{nj}\phi_{ni}^*. \tag{A.13}$$

Comparing Eqs. (A.6) and (A.13) we find that $\boldsymbol{\sigma}$ and $\boldsymbol{A}$ share the same eigenbasis, where their diagonal representations are:

$$A_{ij} = \delta_{ij}\varepsilon_i\,;\; \sigma_{ij} = \delta_{ij}\frac{1}{e^{\beta\varepsilon_i}+1}. \tag{A.14}$$

These can be now used to obtain an expression for the remaining trace appearing in Eq. (A.4) for the entropy:

$$\langle\widehat{\boldsymbol{A}}\rangle = \text{Tr}(\boldsymbol{\rho}\widehat{\boldsymbol{A}}) = \text{Tr}\left(\boldsymbol{\rho}\sum_{i,j}A_{ij}c_i^\dagger c_j\right) = \sum_{i,j}A_{ij}\text{Tr}(\boldsymbol{\rho}c_i^\dagger c_j) = \sum_{i,j}A_{ij}\sigma_{ji} = \text{Tr}(\boldsymbol{A}\boldsymbol{\sigma}). \tag{A.15}$$

Calculating the last trace in the diagonal basis of $\boldsymbol{A}$ and $\boldsymbol{\sigma}$ yields:

$$\text{Tr}(\boldsymbol{\rho}\widehat{\boldsymbol{A}}) = \sum_{i,j}(\delta_{ij}\varepsilon_i)\left(\delta_{ij}\frac{1}{e^{\beta\varepsilon_i}+1}\right) = \sum_i \frac{\varepsilon_i}{e^{\beta\varepsilon_i}+1}. \tag{A.16}$$

Substituting (A.8) and (A.16) into (A.4) gives:

$$S = k_B \sum_k \left[\ln(1+e^{-\beta\varepsilon_k}) + \frac{\beta\varepsilon_k}{e^{\beta\varepsilon_k}+1}\right]. \tag{A.17}$$

Finally, writing Eq. (A.17) in terms of the eigenvalues of the single-electron density matrix (Eq. (A.14)), $s_k = (e^{\beta\varepsilon_k}+1)^{-1}$, yields Eq. (70) of the main text:

$$S = -k_B \sum_k (1-s_k)\ln(1-s_k) + s_k \ln(s_k). \tag{A.18}$$

## Appendix B: Proof that $\langle d|\{\sigma,V\}|d\rangle = \sum_l \langle l|\{\sigma,V\}|l\rangle$

The expectation value of the total electronic energy of the entire (dot-lead) system is given by:

$$E_{sys} = tr(\sigma H). \tag{B.1}$$

We symmetrize this expression to ensure that all partial traces remain real-valued, and write is as a sum of the dot $(E_d)$ and lead $(E_L)$ components:

$$E_{sys} = \frac{1}{2}tr(\sigma H + H\sigma) = \frac{1}{2}tr_d(\sigma H + H\sigma) + \frac{1}{2}tr_L(\sigma H + H\sigma) = E_d + E_L, \tag{B.2}$$

where we have used the cyclic property of the full trace operator and defined:

$$\begin{cases} E_d \equiv \frac{1}{2}\langle d|\sigma H + H\sigma|d\rangle \\ E_L \equiv \frac{1}{2}\sum_l \langle l|\sigma H + H\sigma|l\rangle \end{cases}. \tag{B.3}$$

We can divide the Hamiltonian into its diagonal $(H_0)$ and off-diagonal $(V)$ components:

$$H = \underbrace{\varepsilon_d(t)|d\rangle\langle d| + \sum_l \varepsilon_l |l\rangle\langle l|}_{H_0} + \underbrace{\sum_l v_l(|l\rangle\langle d| + |d\rangle\langle l|)}_{V}, \tag{B.4}$$

where the former represents the isolated dot and lead eigenstates, and the latter their mutual coupling, whose matrix elements are assumed to be real for simplicity.

We may now write $E_d$ as:

$$E_d = \frac{1}{2}tr_d[\sigma(H_0 + V) + (H_0 + V)\sigma] = \frac{1}{2}tr_d(\{\sigma,H_0\}) + \frac{1}{2}tr_d(\{\sigma,V\}). \tag{B.5}$$

The first term in Eq. (B.5) yields:

$$\frac{1}{2}tr_d(\sigma H_0 + H_0 \sigma)$$

$$= \frac{1}{2}\left\langle d\left| \underbrace{(\sigma_d|d\rangle\langle d| + \sum_l \sigma_{dl}|d\rangle\langle l| + \sum_l \sigma_{ld}|l\rangle\langle d| + \sum_{ll'}\sigma_{ll'}|l\rangle\langle l'|)}_{\sigma} \underbrace{(\varepsilon_d(t)|d\rangle\langle d| + \sum_{l''}\varepsilon_{l''}|l''\rangle\langle l''|)}_{H_0} \right| d \right\rangle$$

$$+ \frac{1}{2}\left\langle d\left| \underbrace{(\varepsilon_d(t)|d\rangle\langle d| + \sum_{l''}\varepsilon_{l''}|l''\rangle\langle l''|)}_{H_0} \underbrace{(\sigma_d|d\rangle\langle d| + \sum_l \sigma_{dl}|d\rangle\langle l| + \sum_l \sigma_{ld}|l\rangle\langle d| + \sum_{ll'}\sigma_{ll'}|l\rangle\langle l'|)}_{\sigma} \right| d \right\rangle$$

$$= \frac{1}{2}\sigma_d \varepsilon_d(t) + \frac{1}{2}\varepsilon_d(t)\sigma_d = \varepsilon_d(t)\sigma_d. \tag{B.6}$$

The second term in Eq. (B.5) yields:

$$\frac{1}{2}tr_d(\boldsymbol{\sigma V} + \boldsymbol{V\sigma})$$

$$= \frac{1}{2}\left\langle d \left| \underbrace{(\sigma_d|d\rangle\langle d| + \sum_l \sigma_{dl}|d\rangle\langle l| + \sum_l \sigma_{ld}|l\rangle\langle d| + \sum_{ll'} \sigma_{ll'}|l\rangle\langle l'|)}_{\sigma} \underbrace{\sum_{l''} v_{l''}(|l''\rangle\langle d| + |d\rangle\langle l''|)}_{V} \right| d \right\rangle +$$

$$\frac{1}{2}\left\langle d \left| \underbrace{\sum_{l''} v_{l''}(|l''\rangle\langle d| + |d\rangle\langle l''|)}_{V} \underbrace{(\sigma_d|d\rangle\langle d| + \sum_l \sigma_{dl}|d\rangle\langle l| + \sum_l \sigma_{ld}|l\rangle\langle d| + \sum_{ll'} \sigma_{ll'}|l\rangle\langle l'|)}_{\sigma} \right| d \right\rangle =$$

$$\frac{1}{2}\langle d|\sum_l \sigma_{dl}|d\rangle\langle l|\sum_{l''} v_{l''}|l''\rangle\langle d||d\rangle + \frac{1}{2}\langle d|\sum_{l''} v_{l''}|d\rangle\langle l''|\sum_l \sigma_{ld}|l\rangle\langle d||d\rangle = \frac{1}{2}\sum_l v_l(\sigma_{dl} + \sigma_{ld}). \quad \text{(B.7)}$$

Therefore, we may write:

$$E_d = \varepsilon_d(t)\sigma_d + \frac{1}{2}\sum_l v_l(\sigma_{dl} + \sigma_{ld}). \quad \text{(B.8)}$$

Similarly, $E_L$ can be written as:

$$E_L = \frac{1}{2}tr_L[\boldsymbol{\sigma}(\boldsymbol{H_0} + \boldsymbol{V}) + (\boldsymbol{H_0} + \boldsymbol{V})\boldsymbol{\sigma}] = \frac{1}{2}tr_L(\{\boldsymbol{\sigma}, \boldsymbol{H_0}\}) + \frac{1}{2}tr_L(\{\boldsymbol{\sigma}, \boldsymbol{V}\}). \quad \text{(B.9)}$$

The first term in Eq. (B.9) yields:

$$\frac{1}{2}tr_L(\boldsymbol{\sigma H_0} + \boldsymbol{H_0 \sigma})$$

$$= \frac{1}{2}\sum_{l'''}\left\langle l''' \left| \underbrace{(\sigma_d|d\rangle\langle d| + \sum_l \sigma_{dl}|d\rangle\langle l| + \sum_l \sigma_{ld}|l\rangle\langle d| + \sum_{ll'} \sigma_{ll'}|l\rangle\langle l'|)}_{\sigma} \underbrace{(\varepsilon_d(t)|d\rangle\langle d| + \sum_{l''} \varepsilon_{l''}|l''\rangle\langle l''|)}_{H_0} \right| l''' \right\rangle$$

$$+ \frac{1}{2}\sum_{l'''}\left\langle l''' \left| \underbrace{(\varepsilon_d(t)|d\rangle\langle d| + \sum_{l''} \varepsilon_{l''}|l''\rangle\langle l''|)}_{H_0} \underbrace{(\sigma_d|d\rangle\langle d| + \sum_l \sigma_{dl}|d\rangle\langle l| + \sum_l \sigma_{ld}|l\rangle\langle d| + \sum_{ll'} \sigma_{ll'}|l\rangle\langle l'|)}_{\sigma} \right| l''' \right\rangle$$

$$= \frac{1}{2}\sum_{l'''}\langle l'''|\sum_{ll'}\sigma_{ll'}|l\rangle\langle l'|\sum_{l''}\varepsilon_{l''}|l''\rangle\langle l''||l'''\rangle + \frac{1}{2}\sum_{l'''}\langle l'''|\sum_{l''}\varepsilon_{l''}|l''\rangle\langle l''|\sum_{ll'}\sigma_{ll'}|l\rangle\langle l'||l'''\rangle$$

$$= \frac{1}{2}\sum_l \sigma_{ll}\varepsilon_l + \frac{1}{2}\sum_l \varepsilon_l \sigma_{ll} = \sum_l \varepsilon_l \sigma_{ll}. \quad \text{(B.10)}$$

The second term in Eq. (B.9) yields:

$$\frac{1}{2}tr_L(\boldsymbol{\sigma V} + \boldsymbol{V\sigma}) =$$

$$\frac{1}{2}\sum_{l'''}\left\langle l''' \left| \underbrace{(\sigma_d|d\rangle\langle d| + \sum_l \sigma_{dl}|d\rangle\langle l| + \sum_l \sigma_{ld}|l\rangle\langle d| + \sum_{ll'} \sigma_{ll'}|l\rangle\langle l'|)}_{\sigma} \underbrace{\sum_{l''} v_{l''}(|l''\rangle\langle d| + |d\rangle\langle l''|)}_{V} \right| l''' \right\rangle +$$

$$\frac{1}{2}\sum_{l'''}\left\langle l'''\left|\underbrace{\sum_{l''} v_{l''}(|l''\rangle\langle d| + |d\rangle\langle l''|)}_{V}\underbrace{(\sigma_d|d\rangle\langle d| + \sum_l \sigma_{dl}|d\rangle\langle l| + \sum_l \sigma_{ld}|l\rangle\langle d| + \sum_{ll'} \sigma_{ll'}|l\rangle\langle l'|)}_{\sigma}\right|l'''\right\rangle =$$

$$\frac{1}{2}\sum_{l'''}\langle l'''|\sum_l \sigma_{ld}|l\rangle\langle d|\sum_{l''} v_{l''}|d\rangle\langle l''||l'''\rangle + \frac{1}{2}\sum_{l'''}\langle l'''|\sum_{l''} v_{l''}(|l''\rangle\langle d|)\sum_l \sigma_{dl}|d\rangle\langle l||l'''\rangle =$$

$$\frac{1}{2}\sum_l v_l(\sigma_{ld} + \sigma_{dl}) \tag{B.11}$$

Therefore, we may write:

$$E_L = \sum_l \varepsilon_l \sigma_{ll} + \frac{1}{2}\sum_l v_l(\sigma_{ld} + \sigma_{dl}). \tag{B.12}$$

Comparing Eqs. (B.7) and (B.11) we can see that $tr_L(\boldsymbol{\sigma V} + \boldsymbol{V\sigma}) = tr_d(\boldsymbol{\sigma V} + \boldsymbol{V\sigma})$. Furthermore, from Eqs. (B.8) and (B.12) it becomes clear that the coupling contribution equally splits between the dot and lead energy terms.

# A Numerical Approach to Non-Equilibrium Quantum Thermodynamics: Non-Perturbative Treatment of the Driven Resonant Level Model based on the Driven Liouville von-Neumann Formalism Supporting Information

Annabelle Oz,[1,2] Oded Hod,[1,2] and Abraham Nitzan[1,2,3]

[1] *Department of Physical Chemistry, School of Chemistry, The Raymond and Beverly Sackler Faulty of Exact Sciences, Tel Aviv University, Tel Aviv, IL 6997801*

[2] *The Sackler Center for Computational Molecular and Materials Science, Tel Aviv University, Tel Aviv, IL 6997801*

[3] *Department of Chemistry, University of Pennsylvania, Philadelphia, PA, USA 19103*

The following items appear in this supporting information:

1. DLvN driving rate ($\Gamma$) Sensitivity Check.
2. Finite Bandwidth Lead Model Sensitivity Check.
3. Lead Density of States Sensitivity Check.
4. Evaluation of the Lead Contribution to Quasi-Static Observable Variations.

## 1. DLvN Driving Rate ($\Gamma$) Sensitivity Check

The results presented in the main text were obtained using a finite lead model of $N_L = 100$ levels and a bandwidth of 10 (in units of $\hbar\gamma$). Correspondingly, the DLvN driving rate was set to broaden the lead levels according to their spacing such that $\hbar\Gamma = \Delta\varepsilon = 0.1$. To verify that our results are sufficiently insensitive to this driving rate choice we repeated the excess dot occupation and energy contribution calculations using $\dot{\varepsilon}_d = 0.1$ (in units of $\hbar\gamma^2$) for a lead model size of $N_L = 200$, of bandwidth 10, electronic thermal energy of $K_B T = 0.5$, dot level shift rate of $\dot{\varepsilon}_d = 0.1$, and three values of the DLvN driving rate: $\Gamma = \Delta\varepsilon = 0.05$, $\Gamma = 2\Delta\varepsilon = 0.10$, and $\Gamma = 3\Delta\varepsilon = 0.15$. The results presented in Figs. S1-S2 demonstrate that our numerical calculations are insensitive to the choice of $\Gamma$ within the range of values considered.

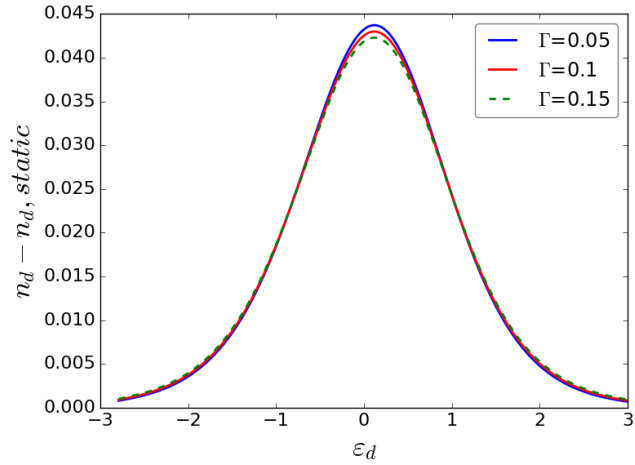

Figure S1: Excess dot occupation as a function of $\varepsilon_d$ for a DLvN driving rate of $\Gamma = 0.05$ (Full blue line), $\Gamma = 0.10$ (Full red line) and $\Gamma = 0.15$ (Dashed green line). In all calculations $\dot{\varepsilon}_d = 0.1$, $K_B T = 0.5$, the lead bandwidth is 10, and the lead size is $N_L = 200$.

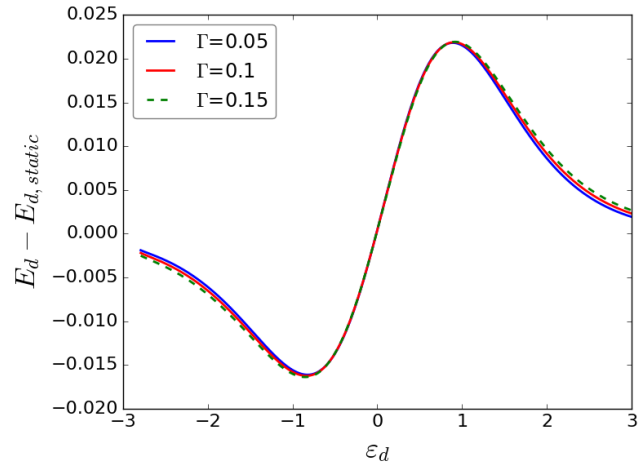

Figure 11S2: Excess dot energy contribution as a function of $\varepsilon_d$ for a DLvN driving rate of $\Gamma = 0.05$ (Full blue line), $\Gamma =$

0.10 (Full red line) and $\Gamma = 0.15$ (Dashed green line). In all calculations $\dot{\varepsilon}_d = 0.1$, $K_B T = 0.5$, the lead bandwidth is 10, and the lead size is $N_L = 200$.

## 2. Finite Bandwidth Lead Model Sensitivity Check

The results presented in the main text were obtained using a finite lead model size of 100 levels spanning a bandwidth $10\,\hbar\gamma$. To verify that our results are sufficiently insensitive to this bandwidth choice we repeated the quasi-static dot occupation, entropy, and heat calculations using a lead bandwidth of $20\hbar\gamma$. To keep the density of lead states constant we also doubled the number of finite lead model levels to 200. The rest of the parameters were the same as those used for the calculations presented in the main text ($\gamma = 1$, $\Gamma = \Delta\varepsilon = 0.1$, and $k_B T = 0.5$). The results presented in Figs. S3-S5 demonstrate that our numerical calculations, which (as discussed in the main text) are in good agreement with the analytical WBL results, are fairly insensitive to the choice of lead bandwidth. Furthermore, the minor changes associated with doubling the bandwidth of the finite lead model make the agreement between the numerical calculation and the WBL expression even better. Therefore, we may conclude that our numerical results using the finite lead model are converged to the WBL case. We note again that our numerical calculations are not limited to the wide-band approximation and the choice to approach this limit here is made deliberately for the purpose of comparison with the analytical expressions.

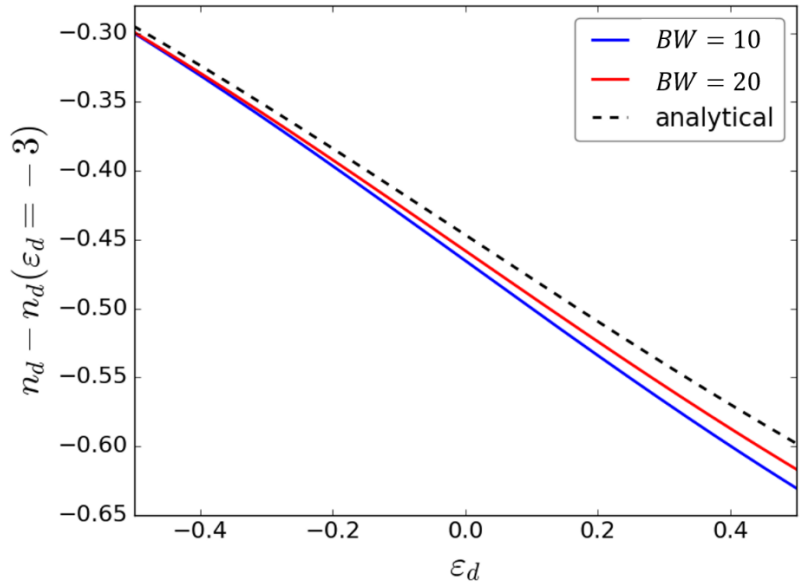

Figure S3: Quasi-static dot occupation measured relative to its values at $\varepsilon_{d1} = -3$, plotted against the dot energy $\varepsilon_d$ for lead model bandwidth of $10\hbar\gamma$ (full blue line) and $20\hbar\gamma$ (full red line) compared to the analytical WBL results (dashed black line). The full blue line and the dashed black line are the same as those appearing in Fig. 2a of the main text.

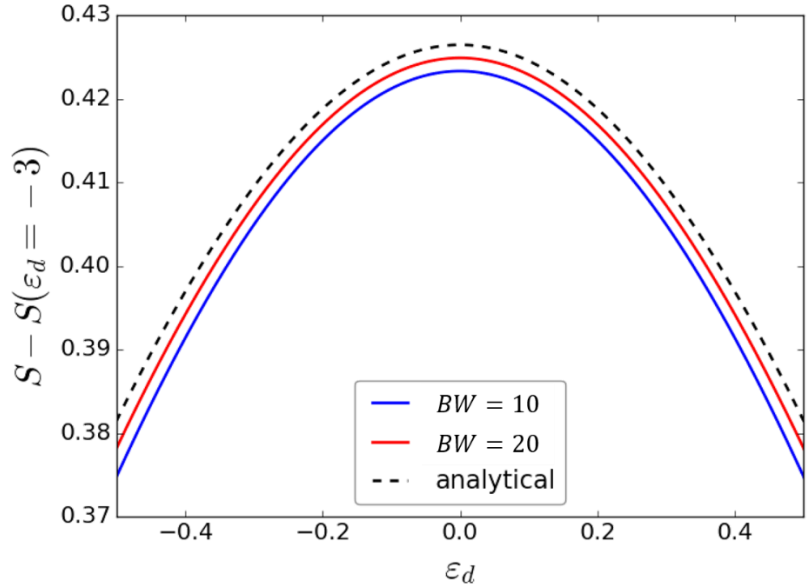

Figure S4 Entropy (calculate via Eq. (22) of the main text) measured relative to its value at $\varepsilon_{d1} = -3$, plotted against the dot energy $\varepsilon_d$ for lead model bandwidth of $10\hbar\gamma$ (full blue line) and $20\hbar\gamma$ (full red line) compared to the analytical WBL results (dashed black line). The full blue line and the dashed black line are the same as those appearing in Fig. 2c of the main text.

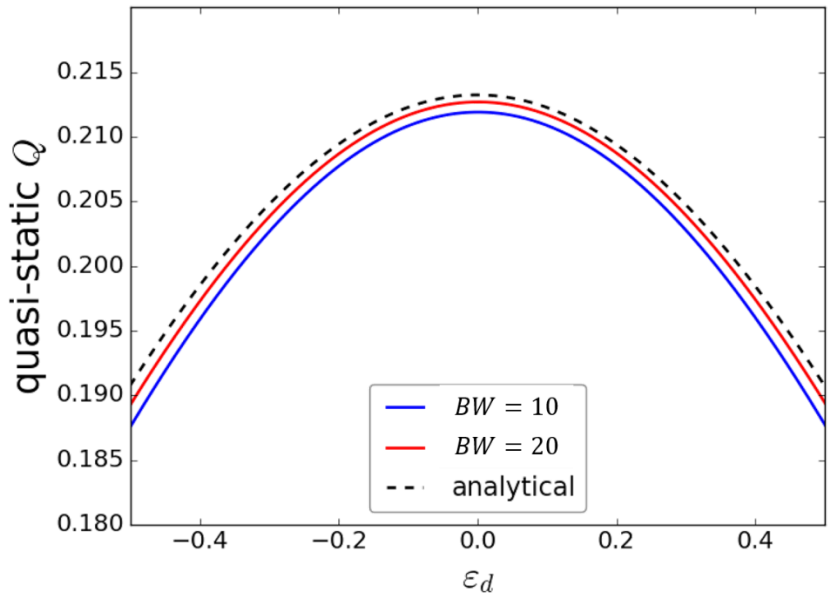

Figure S5: Heat measured relative to its value at $\varepsilon_{d1} = -3$, plotted against the dot energy $\varepsilon_d$ for lead model bandwidth of $10\hbar\gamma$ (full blue line) and $20\hbar\gamma$ (full red line) compared to the analytical WBL results (dashed black line). The full blue line and the dashed black line are the same as those appearing in Fig. 3b of the main text.

## 3. Lead Density of States Sensitivity Check

The results presented in the main text were obtained using a finite lead model of $N_L = 100$ levels and a bandwidth of 10 (in units of $\hbar\gamma$). yielding a lead inter level spacing of $\Delta\varepsilon = 0.1$ (in units of $\hbar\gamma$). To verify that our results are converged with respect to the density of lead states we repeated the excess dot occupation and energy contribution calculations using $\dot{\varepsilon}_d = 0.1$ (in units $\hbar\gamma^2$) for a lead model size of $N_L = 200$ at the same lead bandwidth yielding a lead inter level spacing of $\Delta\varepsilon = 0.05$. The calculations were performed at a dot level shift rate of $\dot{\varepsilon}_d = 0.1$ and bath electronic thermal energy of $K_B T = 0.5$. The DLvN driving rate was set to broaden the lead levels according to their spacing such that $\hbar\Gamma = \Delta\varepsilon$. The results presented in Figs. S6-S7 demonstrate that our numerical calculations, are well converged with respect to the density of finite lead model states.

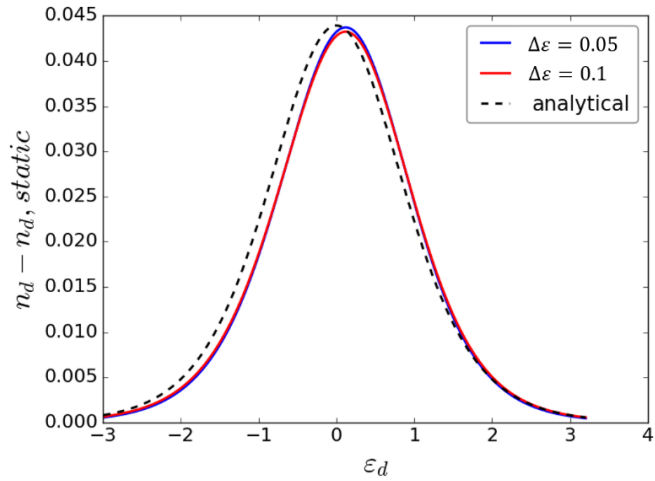

Figure S6: Excess dot occupation as a function of $\varepsilon_d$ for lead inter-level spacing of $\Delta\varepsilon = 0.1$ (full red line) and $\Delta\varepsilon = 0.05$ (full blue line). In both cases we take $\dot{\varepsilon}_d = 0.1$, $K_B T = 0.5$, lead bandwidth of 10, and $\Gamma = \Delta\varepsilon$.

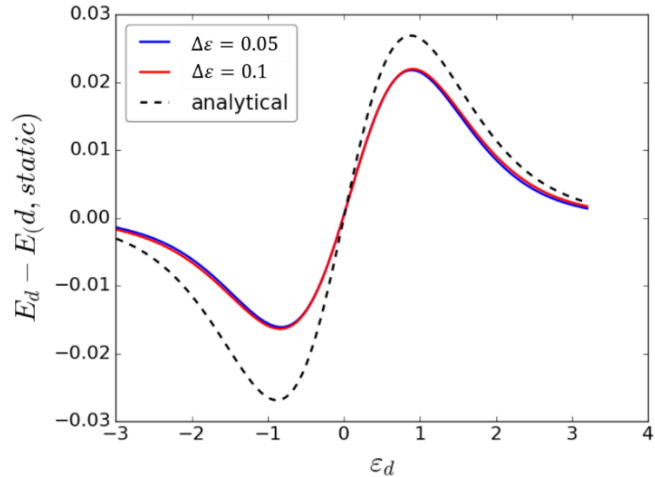

Figure S7: Excess dot energy contribution as a function of $\varepsilon_d$ for lead inter-level spacing of $\Delta\varepsilon = 0.1$ (full red line) and $\Delta\varepsilon = 0.05$ (full blue line). In both cases we take $\dot{\varepsilon}_d = 0.1$, $K_B T = 0.5$, lead bandwidth of 10, and $\Gamma = \Delta\varepsilon$.

## 4. Evaluation of the Lead Contribution to Quasi-Static Observable Variations

In the main text, when comparing the results of our numerical calculations to the WBL analytical expressions, we have focused on the dot's contribution to the variations in particle number and energy. Indeed, in the analytical treatment the lead (representing the entire bath) is assumed to be constantly at equilibrium regardless of the dot dynamics. Hence, all variations occur at the dot itself and its interface with the lead. On the contrary, in our numerical treatment the coupled dot-lead dynamics is considered explicitly. Since the lead is driven towards equilibrium at a finite rate, its dynamical state in any given instance only approximates the desired equilibrium.

To test how much this influences the comparison with the WBL analytical results we have repeated the equilibrium numerical calculations comparing the variations in the number of particles and energy of the entire dot-lead system compared to the dot contributions alone. To this end, we assigned Fermi-Dirac occupations to the eigenstates of the Hamiltonian of the dot-lead system at different dot level positions and compared the variations in the total number of electrons and energy to their projection on the dot site. The small deviations reflected in Figs. S8 and S9 indicate that the lead's deviation from equilibrium, at least under quasi-static conditions, is minor and that most of the variations occur at the dot. Nevertheless, the total dot-lead observables provide a somewhat better fit to the WBL analytical results than the dot contribution alone.

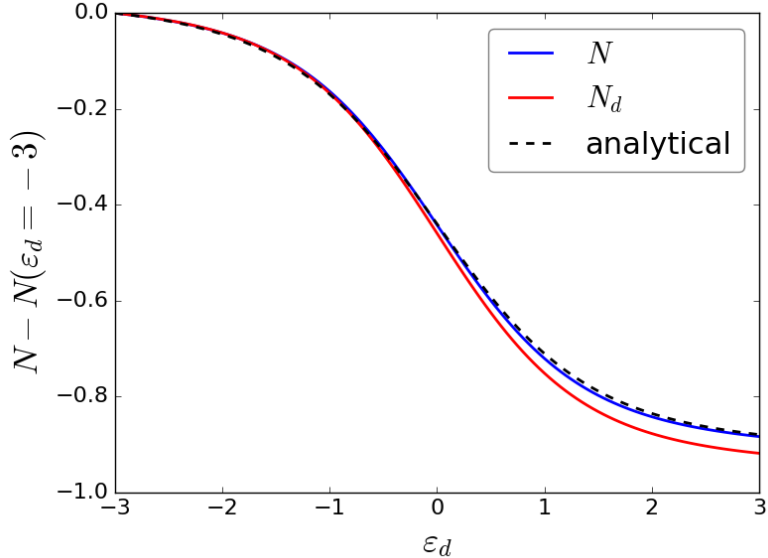

Figure S8: Equilibrium dot (full blue line) and dot-lead (full red line) occupations measured relative to their values at $\varepsilon_{d1} = -3$, plotted against the dot energy $\varepsilon_d$. Results obtained via the analytical WBL expression are presented by the dashed-

black line. The results were obtained at an electronic temperature corresponding to $K_B T = 0.5$ (in units of $\hbar\gamma$) using a finite lead model of $N_L = 100$ levels and a bandwidth of 10. Correspondingly, the DLvN driving rate was set to broaden the lead levels according to their spacing such that $\Gamma = \Delta\varepsilon = 0.1$.

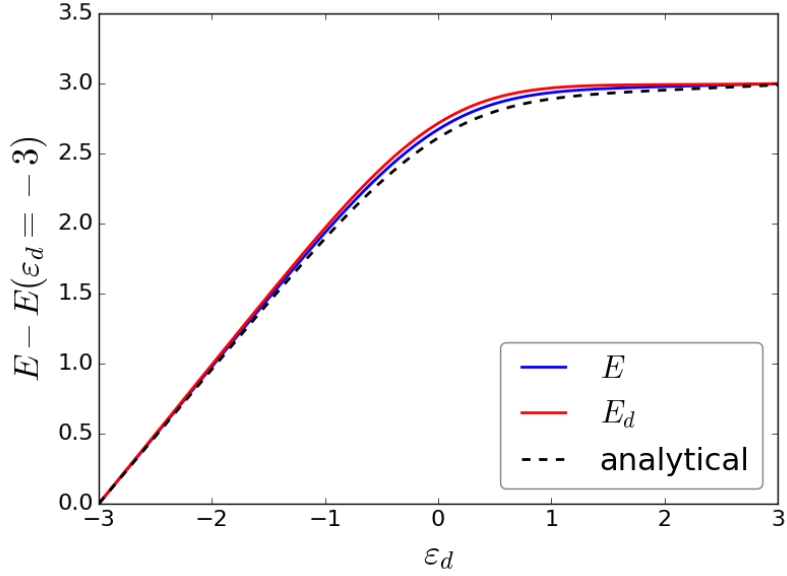

Figure S9: Equilibrium dot (full blue line) and dot-lead (full red line) energies measured relative to their values at $\varepsilon_{d1} = -3$, plotted against the dot energy $\varepsilon_d$. Results obtained via the analytical WBL expression are presented by the dashed-black line. The results were obtained at an electronic temperature corresponding to $K_B T = 0.5$ (in units of $\hbar\gamma$) using a finite lead model of $N_L = 100$ levels and a bandwidth of 10. Correspondingly, the DLvN driving rate was set to broaden the lead levels according to their spacing such that $\Gamma = \Delta\varepsilon = 0.1$.